\documentclass[prd,superscriptaddress,preprintnumbers,twocolumn,showpacs,floatfix]{revtex4}

\usepackage{graphicx}
\usepackage{amsmath}
\usepackage{amssymb}
\usepackage{color}
\usepackage{bm}

\newcommand{\bp}{\mathbf{p}}
\newcommand{\up}{\hat{\mathbf{p}}}
\newcommand{\uP}{\hat{\mathbf{P}}}
\newcommand{\hp}{\hat{p}}

\begin{document}
\preprint{TUM-EFT 4/09}

\title{Non-leptonic weak processes in spin-one color superconducting quark matter}

\author{Xinyang Wang}
\affiliation{Department of Physics, Arizona State University, Tempe, Arizona 85287, USA}

\author{Hossein Malekzadeh}
\affiliation{Institut f\"ur Theoretische Physik, Westf\"alische Wilhelms-Universit\"at M\"unster, D-48149 M\"unster, Germany}
\affiliation{Physik Department, Technische Universit\"at M\"unchen, D-85748 Garching, Germany}

\author{Igor A. Shovkovy}
\email{igor.shovkovy@asu.edu}
\affiliation{Department of Physics, Arizona State University, Tempe, Arizona 85287, USA}
\affiliation{Department of Applied Sciences and Mathematics, Arizona State University, Mesa, Arizona 85212, USA}

\begin{abstract}
The non-leptonic weak processes $s+u\to u+d$ and $u+d \to s+u$ are known to dominate the 
dissipation mechanism responsible for the viscosity of strange quark matter in its normal 
phase. The rates of such processes remain unknown for many color superconducting phases 
of quark matter. In this paper, we partially fill up the gap by calculating the difference 
of the rates of the two non-leptonic weak processes in four transverse spin-one color 
superconducting phases of quark matter (slightly) out of $\beta$-equilibrium. The four 
phases studied are the color-spin locked phase, the polar phase, the planar phase and 
the {\em A}-phase. In the limit of vanishing color superconducting gap, we reproduce 
the known results in the normal phase. In the general case, the rates are suppressed 
relative to the normal phase. The degree of the suppression is determined by the 
structure of the gap function in momentum space, which in turn is determined by 
the pairing pattern of quarks. At low temperatures, the rate is dominated by the ungapped 
modes. In this limit, the strongest suppression of the rate occurs in the color-spin-locked 
phase, and the weakest is in the polar phase and the {\em A}-phase. 
\end{abstract}

\date{\today}


\maketitle

\section{Introduction}

The interior of neutron stars is made of very dense baryonic matter. Currently our
knowledge regarding the actual state of such matter is incomplete. One commonly 
accepted hypotheses is that the densest regions inside neutron stars are made of quark 
matter \cite{Collins:1974ky}. Moreover, such quark matter may be a color superconductor
\cite{Barrois:1977xd,Bailin:1984dz}. (For reviews on color superconductivity see for 
example Refs.~\cite{rev1,rev2,rev3,rev4,rev5,Buballa:2003qv,Shovkovy:2004me,Alford:2007xm,Wang:2009xf}.)
From the viewpoint of basic research, it is of fundamental importance to test this 
hypothesis empirically. 

The way to test the idea regarding the presence of quark matter inside stars is 
to make predictions regarding physics processes that affect observable features 
of stars and then test them against the stellar data. One class of physics 
properties that are substantially modified by the presence of color superconducting 
quark matter is related to the rates of weak processes. Such processes, for example, 
affect the cooling rates \cite{iwamoto} and the suppression of the rotational 
(r-mode) instabilities \cite{Andersson:1997xt} in stars. The latter in particular 
is determined by the viscous properties of dense matter \cite{Madsen:1998qb}.

Theoretically, the ground state of baryonic matter at very high density corresponds to 
the color-flavor-locked (CFL) phase of quark matter \cite{Alford:1998mk}. In this phase, 
quarks of all three colors and all three flavors participate in spin-zero Cooper pairing on 
equal footing. The rates of the weak processes and some of their effects on the physical 
properties of the CFL phase of quark matter have been discussed in
Refs.~\cite{Jaikumar:2002vg,Reddy:2002xc,Alford:2007rw,Jaikumar:2008kh,Alford:2009jm}.

With decreasing the density, the CFL phase should break up. This 
is due to the disruptive effects of a large difference between the masses of the strange 
quark and the light (up and down) quarks \cite{Alford:2002kj}. Such a difference leads 
to a mismatch between the Fermi momenta of quarks and, therefore, spoils the ``democratic" 
pairing of the CFL phase. When the CFL phase breaks up, another type of spin-zero color 
superconductivity, the so-called two-flavor color superconducting (2SC) phase~\cite{cs}, 
can still be possible. In the 2SC phase, strange quarks do not participate in pairing. 
Also, up and down quarks of one color remain unpaired. Some weak processes in the 2SC 
phase and their effects on the physical properties have been studied in 
Refs.~\cite{Alford:2006gy,Jaikumar:2005hy}.

It is important to mention that matter inside stars is neutral (at least on average) and 
in $\beta$-equilibrium. Enforcing these two conditions affects the pairing between quarks 
and may disrupt the usual formation of cross-flavor spin-zero Cooper pairs \cite{Alford:2002kj}. 
In this case, the ground state of matter can be in other forms, for example, such as stable 
variants of crystalline \cite{Alford:2000ze}, gapless \cite{Shovkovy:2003uu,Alford:2003fq}, 
or other exotic phases \cite{pioncond,gluonic}. When spin-zero pairing cannot occur, the 
ground state can be in one of the spin-one color superconducting phases, in which same 
flavor quarks combine to form Cooper pairs 
\cite{Schafer:2000tw,Schmitt:2004et,spin-1,spin-1-Meissner,Brauner:2009df}. 

Compared to the spin-zero case, the energy gap in spin-one color 
superconductors is likely to be about two orders of magnitude smaller. This means 
that the actual value of the gap may be somewhere in the range from $0.01\,\mbox{MeV}$ 
to $1\,\mbox{MeV}$. It appears that even such relatively small gaps can substantially 
affect the cooling rate of a quark star~\cite{Schmitt:2005wg}. By the same token, such 
gaps can strongly modify the rates of the non-leptonic weak processes and, thus, 
affect the viscosity of stellar quark matter.

\begin{figure}[t]
\includegraphics[width=0.35\textwidth]{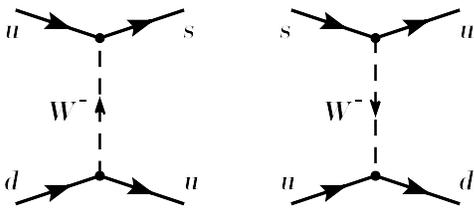}
\caption{Feynman diagrams of the non-leptonic weak processes}
\label{weak-proc}
\end{figure}

The bulk viscosity in the normal phase of three-flavor quark matter is usually dominated 
by the non-leptonic weak processes~\cite{WangLu,Sawyer,madsen,Xiaoping,Zheng,Dong:2007mb}. 
(The corresponding processes are diagrammatically shown in Fig.~\ref{weak-proc}.) 
It was argued in Ref.~\cite{Sa'd:2007ud}, however, that the interplay between the Urca 
and non-leptonic processes may be rather involved even in the normal phase of quark 
matter. Indeed, because of the resonance-like dynamics responsible for the bulk viscosity 
and because of a subtle interference between the two types of the weak processes, a 
larger rate of the non-leptonic processes may not automatically mean its dominant role.
In fact, it was shown that the contributions of the two types of weak processes are not 
separable and that, at low frequencies relevant for some pulsars, taking into account the 
Urca processes may substantially modify the result \cite{Sa'd:2007ud}. 

Currently it remains unknown how a similar interplay between the two types of 
weak processes is realized in spin-one color superconducting phases. Primarily, this is 
because the rates of the non-leptonic weak processes in the corresponding phases have 
not been calculated. (Note that the rates of the Urca processes in several spin-one 
color superconducting phases were obtained in Refs.~\cite{Schmitt:2005wg,Sa'd:2006qv}.)
The purpose of this paper is to study the corresponding non-leptonic rates.  

The rest of the paper is organized as follows. The derivation of a general expression 
for the non-leptonic rate, based on the Kadanoff-Baym formalism \cite{KB}, is presented in 
the next section. The structure of the quark propagators in spin-one color superconducting
phases is described in Subsec.~\ref{QuarkPropagators}. This is used in Subsec.~\ref{PolarizationTensor}
to derive the imaginary part of the $W$-boson polarization tensor, which is the key ingredient 
in the expression for the rate. The net rate of the $d$-quark production (i.e., the difference 
of the rates of $s+u\to u+d$ and $u+d \to s+u$) 
in the case of a small deviation from chemical equilibrium is obtained in Sec.~\ref{Calculation}. 
There we also present the numerical results for each of the following spin-one color 
superconducting phases: the CSL phase (Subsec.~\ref{CSLPhase}), the polar 
phase (Subsec.~\ref{PolarPhase}), the A-phase (Subsec.~\ref{Aphase}), and the planar 
phase (Subsec.~\ref{PlanarPhase}). In Sec.~\ref{Discussion}, we discuss 
the main results and their physical meaning. Two Appendices at the end of the paper 
contain some details, used in the derivation of the rate.

\section{Formalism}
\label{Formalism}

\begin{figure}[t]
\begin{center}
\includegraphics[width=0.4\textwidth]{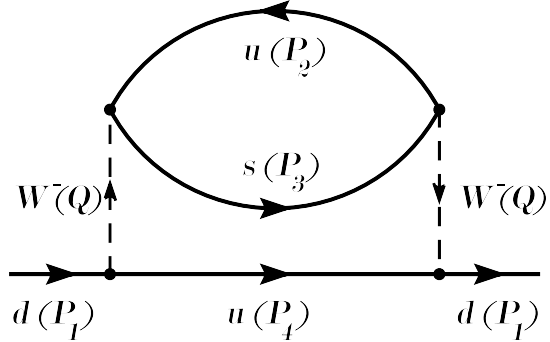}
\caption{Feynman diagram for the $d$-quark self-energy. The particle 
four-momenta are shown in parenthesis next to the particle names.}
\label{self}
\end{center}
\end{figure}

In order to calculate the rates of the non-leptonic processes, we use the same approach 
as in Refs.~\cite{Schmitt:2005wg,Alford:2006gy,Sa'd:2006qv,Sa'd:2007ud,Sedrakian:2000kc}. 
It is based on the Kadanoff-Baym formalism~\cite{KB}. The starting point of the
analysis is the general Kadanoff-Baym equation for the Green functions
(propagators) of the down (or strange) quarks. After applying the conventional 
gradient expansion close to equilibrium, we derive the following kinetic equation 
for the $d$-quark Green function:
\begin{eqnarray} 
i\frac{\partial}{\partial t}{\rm Tr}[\gamma_0S_d^<(P_1)] &=& -{\rm Tr}[S_d^>(P_1)\Sigma^<(P_1) \nonumber \\ 
&&-\Sigma^>(P_1)S_d^<(P_1)] \, .
\label{KineticEquation}
\end{eqnarray}
Here we denote the quark four-momenta by capital letters, e.g., $P = (p_{0}, {\bf p})$,
where $p_0$ is the energy and ${\bf p}$ is the three-momentum. The structure of the quark Green's 
functions $S^{<}(P_1)$ and $S^{>}(P_1)$ in spin-one color superconducting phases will be discussed 
in the next subsection. To leading order, the quark self-energies $\Sigma^{<}(P_1)$ and 
$\Sigma^{>}(P_1)$ are given by the Feynman diagram in Fig.~\ref{self}. 
This translates into the following explicit expression:
\begin{equation} 
\Sigma^{<,>}(P_1) = \frac{i}{M_W^4}
\int\frac{d^4P_4}{(2\pi)^4}\Gamma_{ud,-}^\mu S_u^{<,>}(P_4)\Gamma_{ud,+}^\nu
\Pi_{\mu\nu}^{>,<}(Q) ,
\label{SelfEnergy}
\end{equation}
where, by definition, $M_W$  and $Q=P_1-P_4$ are the mass and the four-momentum of the $W$-boson, 
respectively. (Note that the large hierarchy between the $W$-boson mass and a typical momentum 
transfer $Q\lesssim 1$~MeV justifies the approximation in which the $W$-boson propagator is 
replaced by $1/M_W^2$.) As seen from the diagram in Fig.~\ref{self}, the expression for the 
polarization tensor of the $W$-boson is given by
\begin{widetext}
\begin{equation} 
\Pi_{\mu\nu}^{<,>}(Q) = -i \int\frac{d^4P_2}{(2\pi)^4}{\rm Tr}
\Big[\Gamma_{us,+}^\mu \,S_s^{>,<}(P_2+Q)\Gamma_{us,-}^\nu\, S_u^{<,>}(P_2)\Big] \, .
\label{Pi-W}
\end{equation}
In the Nambu-Gorkov notation used here, the explicit form of the (tree-level) vertices for the 
weak processes $d\leftrightarrow u + W^-$ and $s\leftrightarrow u + W^-$ reads \cite{Alford:2006gy}
\begin{equation} 
\Gamma_{ud/us,\pm}^\mu = \frac{e\,V_{ud/us}}{2\sqrt{2}\sin\theta_W}\left(\begin{array}{cc}
\gamma^\mu(1-\gamma^5)\,\tau_{ud/us,\pm} & 0 \\
0 &  -\gamma^\mu(1+\gamma^5)\,\tau_{ud/us,\mp}\end{array}\right) \, .
\label{Gamma-ud-us}
\end{equation}
These are given in terms of the elements of the Cabibbo-Kobayashi-Maskawa matrix $V_{ud}$ and $V_{us}$, 
and the weak mixing angle $\theta_W$. By construction, the $\tau$-matrices operate in flavor space 
($u,d,s$) and have the following form:
\begin{eqnarray}
\tau_{ud,+} \equiv \left(\begin{array}{ccc}   0 & 1 &0 \\ 0 &0&0 \\ 0&0&0 \end{array}\right) ,
\quad \tau_{ud,-} \equiv \left(\begin{array}{ccc}   0 & 0 &0 \\ 1 &0&0 \\ 0&0&0 \end{array}\right) , 
\quad \tau_{us,+} \equiv \left(\begin{array}{ccc}   0 & 0 &1 \\ 0 &0&0 \\ 0&0&0 \end{array}\right),
\quad \tau_{us,-} \equiv \left(\begin{array}{ccc}   0 & 0 &0 \\ 0 &0&0 \\ 1&0&0 \end{array}\right) .
\end{eqnarray}
By making use of Eqs.~(\ref{KineticEquation}) and (\ref{SelfEnergy}), the kinetic equation takes the following form:
\begin{equation}
i\frac{\partial}{\partial t}{\rm Tr}[\gamma_0S_d^<(P_1)] = -\frac{i}{M_W^4}\int\frac{dP_4}{(2\pi)^4}
{\rm Tr}\Big[S_d^>(P_1)\Gamma_{ud,-}^\mu S_u^<(P_4) \Gamma_{ud,+}^\nu \Pi^>_{\mu\nu}(Q) 
- \Gamma_{ud,-}^\mu S_u^>(P_4)\Gamma_{ud,+}^\nu S_d^<(P_1) \Pi^<_{\mu\nu}(Q)\Big] .
\label{KinEq}
\end{equation}
\end{widetext}
The physical meaning of the expression on the left hand side of this equation is the time derivative
of the $d$-quark distribution function. By integrating this over the complete phase space, we obtain 
the net rate of the $d$-quark production:
\begin{equation}
\Gamma_d\equiv-\frac{i}{4}\frac{\partial}{\partial t} \int\frac{dP_1}{(2\pi)^4}{\rm Tr}[\gamma_0 S_d^<(P_1)].
\end{equation}
Then, by making use of the kinetic equation (\ref{KineticEquation}), we derive
\begin{eqnarray}
\Gamma_d &=& \frac{i}{4M_W^4}\int\frac{d^4P_1 }{(2\pi)^4}\int\frac{dP_4}{(2\pi)^4} \nonumber \\ 
&\times& \Big[{\rm Tr}\left(S_d^>(P_1)\Gamma_{ud,-}^\mu S_u^<(P_4) \Gamma_{ud,+}^\nu  \right)\Pi^>_{\mu\nu}(Q) \nonumber \\ 
&&- {\rm Tr}\left(\Gamma_{ud,-}^\mu S_u^>(P_4)\Gamma_{ud,+}^\nu S_d^<(P_1)  \right) \Pi^<_{\mu\nu}(Q) \Big].
\label{rate-Gamma}
\end{eqnarray}
This rate should be non-vanishing only if the rates of the two 
non-leptonic weak processes $u+s\rightarrow d+u$ and $d+u\rightarrow u+s$ differ. In 
$\beta$-equilibrium, in particular, the latter two should be equal and the net rate of the 
$d$-quark production should vanish. The corresponding state of equilibrium in dense quark 
matter is reached when the chemical potentials of all three quark flavors are equal, i.e., 
$\mu_u = \mu_d= \mu_s$. (For simplicity, here it is assumed that all three quark flavors 
are approximately massless and, therefore, that the electrical neutrality of quark matter 
is achieved without the need for the electrons.)

When the system is forced out of equilibrium, e.g., during the density oscillations caused 
by the collective modes of stellar matter, small deviations from $\beta$-equilibrium are 
induced. For our purposes, the corresponding state can be described by the following 
set of the chemical potentials: $\mu_u = \mu_d = \mu $ and $\mu_s = \mu + \delta \mu$, 
where $\delta \mu $ is a small parameter that characterizes the magnitude of the 
departure from the equilibrium state. Out of equilibrium, the net production of $d$-quarks 
may be nonzero. For example, if $\delta \mu >0$ ($\delta \mu <0$) the system has a deficit 
(an excess) of the down quarks and an excess (a deficit) of the strange quarks. Then, 
one of the weak processes, i.e., $u+s\rightarrow d+u$ ($d+u\rightarrow u+s$), will start 
to dominate over the other in order to restore the equilibrium. The net rate of the 
$d$-quark (or equivalently $s$-quark) production characterizes how quickly this happens.

In order to calculate the net rate of $d$-quark production, however, one needs to know 
the explicit structure of the quark propagators in the specific spin-one color 
superconducting phases. The knowledge of the quark propagators is also needed for 
the calculation of the polarization tensors $\Pi^<_{\mu\nu}(Q) $ and $\Pi^>_{\mu\nu}(Q)$. 
These are discussed in the next two subsections.

\subsection{Quark propagator in a spin-one color superconductor}
\label{QuarkPropagators}

In general, Cooper pairs in spin-one color superconducting phases are given by diquarks 
in a color antitriplet (antisymmetric) and spin triplet (symmetric) state. Depending on
a specific color-spin structure, which is determined by the alignments of the antitriplet 
in color space and the triplet in spin (coordinate) space, many inequivalent color 
superconducting phases may form. 

Each phase is unambiguously specified by the structure 
of the gap matrix, which is commonly written in the following form \cite{Schmitt:2004et}:
\begin{equation}
\Phi (P)^{+} = \sum_{e=\pm}  \phi^{e}(P) {\cal M}_{\bf p}\Lambda_{\bf p}^e ,
\label{phi_1sc}
\end{equation}
where $\phi^{e}(P)$ is the gap function. The Dirac matrices $\Lambda_{\bf p}^e\equiv 
(1+e\gamma_0\bm{\gamma}\cdot\up)/2$, with $e=\pm$, are the projectors onto the positive 
and negative energy states. The color structure of $\Phi (P)^{+}$ is determined by 
\begin{equation}
{\cal M}_{\bf p} = \sum_{i,j=1}^{3} J^i \Delta_{ij}
\left[\hat{p}^j\cos\theta +\gamma^j_{\perp}\sin\theta\right].
\label{delta_1sc}
\end{equation}
where $(J^i)^{jk} = -i \epsilon^{ijk}$ are the antisymmetric matrices in color space, 
$\hat{\mathbf{p}}\equiv \mathbf{p}/p $ is the unit vector in the direction of the 
quasiparticle three-momentum $\mathbf{p}$, and
$\gamma^j_{\perp}\equiv \gamma^j-\hat{p}^j (\bm{\gamma}\cdot\hat{\mathbf{p}})$.
The explicit form of the $3\times 3$ matrix $\Delta_{ij}$ and the value of the 
angular parameter $\theta$ determine specific phases of superconducting matter. 
Among them, there is a number of inert and noninert spin-one phases 
\cite{Brauner:2009df}, which are naturally characterized by the continuous and 
discrete symmetries preserved in the ground state. 

In the two special cases, $\theta=0$ and $\theta=\pi/2$, the corresponding phases 
are called longitudinal and transverse, respectively. In this paper we focus on 
the transverse phases ($\theta=\pi/2$), in which only quarks of the opposite 
chiralities pair and which have lower free energies than the longitudinal phases 
\cite{Schmitt:2004et}. To further constrain the large number of possibilities, 
we concentrate only on the following four most popular ones: the color-spin locked 
phase (CSL), the A-phase, the polar phase and the planar phase. 

The structure of the matrices $\Delta_{ij}$ and ${\cal M}_{\bf p}$ for the mentioned four 
phases are quoted in the first two rows of Tab.~\ref{tablephases} (for more details 
see Refs.~\cite{Schmitt:2004et}). In the corresponding ground states, the 
original symmetry SU(3)$_c\times $SO(3)$_J\times $U(1)$_{\rm em}$ of one-flavor 
quark matter breaks down to 
\cite{Schafer:2000tw,Schmitt:2004et,spin-1,spin-1-Meissner,Brauner:2009df}
\begin{eqnarray*}
\begin{array}{lll}
\widetilde{SO}(3)_{J}  & \quad &  \mbox{(CSL)} ,\\
SU(2)_c\times \widetilde{SO}(2)_J \times \tilde{U}(1)_{\rm em} &  \quad &  \mbox{(A-phase)},\\
SU(2)_c\times SO(2)_J \times \tilde{U}(1)_{\rm em} &  \quad &  \mbox{(polar)},\\
\widetilde{SO}(2)_J \times \tilde{U}(1)_{\rm em} &  \quad & \mbox{(planar)},
\end{array}
\end{eqnarray*}
respectively. 

\begin{table*}[t]
\begin{center}
\begin{ruledtabular}
\begin{tabular}[t]{c|cccc}
 & CSL phase & planar phase & polar phase & ${\it A}$-phase \\
\hline
$\Delta_{ij}$ & $\delta_{ij}$ & $\delta_{i1}\delta_{j1}+\delta_{i2}\delta_{j2}$ &
$\delta_{i3}\delta_{j3}$ & $\delta_{i3}(\delta_{j1} + i\,\delta_{j2})$ \\
${\cal M}_{\bf p}$ & ${\bf J}\cdot\bm{\gamma}_{\perp}(\up)$ & $J_1\gamma_{\perp,1}(\up) + J_2\gamma_{\perp,2}(\up)$ &
$J_3\gamma_{\perp,3}(\up)$ & $J_3[\gamma_{\perp,1}(\up) + i\,\gamma_{\perp,2}(\up)]$ \\
$\lambda_{{\bf p},1}\;(n_1)$ & $2\;(8)$ & $1 + \cos^2\theta_{\bf p}\;(8)$ &  $\sin^2\theta_{\bf p} \;(8)$ &
$(1 + |\cos\theta_{\bf p}|)^2\;(4)$ \\
$\lambda_{{\bf p},2}\;(n_2)$ &  $0\;(4)$ &  $0\;(4)$ &  $0\;(4)$ & $(1 - |\cos\theta_{\bf p}|)^2\;(4)$ \\
$\lambda_{{\bf p},3}\;(n_3)$ & --- & --- & --- &  $0 \;(4)$ \\
\end{tabular}
\end{ruledtabular}
\end{center}
\caption{Matrices $\Delta_{ij}$ and ${\cal M}_{\bf p}$ as well as the eigenvalues
$\lambda_{{\bf p},r}$ with the corresponding degeneracies $n_r$ in four
spin-one color superconducting phases. The angle between ${\bf p}$ and
the $z$-axis is denoted by $\theta_{\bf p}$.}
\label{tablephases}
\end{table*}

In spin-one color superconductors, there is no cross-flavor pairing and, 
therefore, the quark propagator is diagonal in flavor space, i.e., 
\begin{equation} 
S(P) = {\rm diag}[S_u(P),S_d(P),S_s(P)].
\end{equation}
The Nambu-Gorkov structure of each flavor-diagonal element is given by
\begin{equation} 
S_f^{<,>}(P) = 
\left(
\begin{array}{cc} G_{f,+}^{<,>}(P) & F_{f,-}^{<,>}(P) \\
                  F_{f,+}^{<,>}(P) & G_{f,-}^{<,>}(P) 
\end{array}
\right)\, , 
\label{prop}
\end{equation}
where $f=u,d,s$.
The normal (diagonal) and anomalous (off-diagonal) components of the Nambu-Gorkov propagator
have the following structure \cite{Schmitt:2004et}:
\begin{eqnarray}
\label{G-diagonal}
G_{f,\pm}^{<,>}(P) &=&\gamma_0 \Lambda_{\bf p}^{\mp}\sum_{r}\mathcal{P}_{{\bf p},r}^{\pm}G_{\pm,r,f}^{<,>}(P)\, ,
\\
\label{F-plus}
F_{f,+}^{<,>}(P)&=&-\gamma_0 {\cal M}_{\bf p}\gamma_0  \sum_{e,r}\mathcal{P}_{{\bf p},r}^{+} \Lambda_{\bf p}^{-e}F_{+,r,f}^{<,>}(P)\, ,
\\
\label{F-minus}
F_{f,-}^{<,>}(P)&=&-{\cal M}_{\bf p}^{\dagger}\sum_{e,r}\mathcal{P}_{{\bf p},r}^{-} \Lambda_{\bf p}^{e}F_{-,r,f}^{<,>}(P)\, .
\end{eqnarray}
Here, $r$ labels different quasiparticle excitations in color-superconducting quark matter. 
The matrices $\mathcal{P}_{{\bf p},r}^{-}$ and $\mathcal{P}_{{\bf p},r}^{+}$ are the projectors onto 
the subspaces spanned by the eigenvectors of ${\cal M}_{\bf p}{\cal M}^{\dagger}_{\bf p}$ and 
$\gamma^0{\cal M}^{\dagger}_{\bf p}{\cal M}_{\bf p}\gamma^0$, respectively. The explicit 
form of the projectors for each phase can be found in Ref.~\cite{Schmitt:2005wg}. It should 
be noted that both matrices ${\cal M}_{\bf p}{\cal M}^{\dagger}_{\bf p}$ and 
$\gamma^0{\cal M}^{\dagger}_{\bf p}{\cal M}_{\bf p}\gamma^0$ have the same set of eigenvalues 
$\lambda_{{\bf p},r}$,
\begin{eqnarray}
{\cal M}_{\bf p}{\cal M}^{\dagger}_{\bf p} &\equiv &
\sum_{r} \lambda_{{\bf p},r}\mathcal{P}_{{\bf p},r}^{-} \,\, , \\
\gamma^0{\cal M}^{\dagger}_{\bf p}{\cal M}_{\bf p}\gamma^0 &\equiv &
\sum_{r} \lambda_{{\bf p},r}\mathcal{P}_{{\bf p},r}^{+} \,\, .
\end{eqnarray}
The list of all eigenvalues as well as their degeneracies are given in the last three rows 
of Tab.~\ref{tablephases}. Each of the eigenvalues determines a quark quasiparticle with 
the following dispersion relation:
\begin{equation}
\epsilon_{\mathbf{p},r, f}=\sqrt{(p-\mu_f)^2+\vert \phi \vert ^2\lambda_{\mathbf{p} ,r,f}}.
\label{qqe}
\end{equation}
The separate components of the propagators in subspaces spanned by the eigenvectors, see 
Eqs.~(\ref{G-diagonal}), (\ref{F-plus}) and (\ref{F-minus}), can be conveniently rewritten 
in terms of the corresponding distribution functions $f(\epsilon_{p,r,f})$ and the Bogoliubov 
coefficients $B_{\mathbf{p},r,f}^\pm$, i.e.,
{\small
\begin{equation}
G_{\pm,r,f}^>(P) =-2\pi i \sum_{e=\pm} B_{\mathbf{p},r,f}^{\pm e} 
f(e\epsilon_{p,r,f})\delta ( p_0\pm\mu_{f}-e\epsilon_{p,r,f}) , 
\label{Ggtr}
\end{equation}
\begin{equation}
G_{\pm,r,f}^<(P) = -2\pi i \sum_{e=\pm} B_{\mathbf{p},r,f}^{\pm e} 
f(-e\epsilon_{p,r,f}) \delta(p_0\pm\mu_{f}-e\epsilon_{p,r,f}) ,
\label{Gless}
\end{equation}
\begin{equation}
F_{\pm,r,f}^>(P) = 2\pi i\,\frac{\phi}{2\epsilon_{p,r,f}} \sum_{e=\pm} 
e f(e\epsilon_{p,r,f}) \delta(p_0\mp\mu_{f}-e\epsilon_{p,r,f}) ,
\label{Fgtr}
\end{equation}
\begin{equation}
F_{\pm,r,f}^<(P) = 2\pi i\,\frac{\phi}{2\epsilon_{p,r,f}} \sum_{e=\pm} 
e f(-e\epsilon_{p,r,f})\delta(p_0\pm\mu_{f}-e\epsilon_{p,r,f}) .
\label{Fless}
\end{equation}
}
The Bogoliubov coefficients and the fermion distribution function are defined as 
follows:
\begin{eqnarray}
B_{\mathbf{p},r,f}^{e} &=& \frac{1}{2}-e\,\frac{p-\mu_f}{2\epsilon_{{\bf p},r,f}},
\\
f(\epsilon)&=&\frac{1}{\exp(\frac{\epsilon}{T})+1}.
\end{eqnarray}
[Note that $f(-\epsilon)=1-f(\epsilon)$.]
The quark propagators in Eq.~(\ref{prop}) can now be used to derive the general 
expressions for $\Pi^<_{\mu\nu}(Q) $ and $\Pi^>_{\mu\nu}(Q)$. This is done in the next
subsection. The results are then used to calculate the rate $\Gamma_d$ in Eq.~(\ref{rate-Gamma}).

\subsection{$W$-boson polarization tensor}
\label{PolarizationTensor}

The $W$-boson polarization tensor is given in terms of the quark propagators in Eq.~(\ref{Pi-W}). 
By taking into account the Nambu-Gorkov and flavor structure of the weak interaction vertices 
in Eq.~(\ref{Gamma-ud-us}), as well as the quark propagator in Eq.~(\ref{prop}), we derive 
\begin{eqnarray}
&& \hspace*{-0.3in}
\Pi_{\mu\nu}^{<,>}(Q) =-\frac{i e^2 V_{us}^2}{8\sin^2\theta_{W}}\int\frac{d^4P_2}{(2\pi)^4}\nonumber\\
&& \times {\rm Tr}\Big[\gamma^{\mu}(1-\gamma^5)G_{s,+}^{>,<}(P_3)\gamma^{\nu}(1-\gamma^5)G_{u,+}^{<,>}(P_2) \nonumber \\ 
&& +\gamma^{\mu}(1+\gamma^5)G_{u,-}^{>,<}(P_3)\gamma^{\nu}(1+\gamma^5)G_{s,-}^{<,>}(P_2) \Big] \, ,
\label{polPi}
\end{eqnarray}
where we introduced the notation $P_3 \equiv P_2 + Q$. Note that the anomalous (off-diagonal) 
elements of the Nambu-Gorkov propagators dropped out from the result. This is the consequence 
of the electric charge conservation. In calculations, this comes about as a result of the 
specific flavor structure of the weak interaction vertices in Eq.~(\ref{Gamma-ud-us}). 
One can further simplify the result for the polarization tensor in Eq.~(\ref{polPi}) 
by noticing that the two terms on the right hand side are equal. From physical viewpoint, 
this is related to the fact that the two terms are the charge-conjugate contributions 
of each other. After taking this into, we arrive at the following expression for the 
polarization tensor:
\begin{eqnarray}
\Pi_{\mu\nu}^{<,>}(Q) &=&  -\frac{i e^2 V_{us}^2}{4\sin ^2\theta_{W}} \int\frac{d^4P_2}{(2\pi)^4}
{\rm Tr} \Big[\gamma^{\mu}(1-\gamma^5) \nonumber\\
& & \times G_{s,+}^{>,<}(P_3)\gamma^{\nu}(1-\gamma^5)G_{u,+}^{<,>}(P_2)\Big] \, .
\end{eqnarray}
Then, by using the explicit structure of the normal components of the $u$- and $s$-quark propagators, 
defined in Eq.~(\ref{G-diagonal}), we obtain
\begin{widetext}
\begin{equation}
\Pi_{\mu\nu}^{<,>}(Q)= -\frac{i e^2 V_{us}^2}{4\sin ^2\theta_{W}} \int\frac{d^4P_2}{(2\pi)^4}
{\rm Tr}\Big[\gamma^{\mu}(1-\gamma^5)\gamma_0 \Lambda _{\mathbf{p}_3}^-\sum _{r_3} \mathcal{P}_{\mathbf{p}_3,r_3}^+
G_{+,r_3,s}^{>,<}(P_3) \gamma^{\nu}(1-\gamma^5)  \gamma_0\Lambda _{\mathbf{p}_2}^{-}
\sum _{r_2} \mathcal{P}_{\mathbf{p}_2,r_2}^+G_{+,r_2,u}^{<,>}(P_2)\Big] \, .
\end{equation}
This can be rewritten in an equivalent form as 
\begin{equation}
\Pi_{\mu\nu}^{<,>} = -\frac{i e^2 V_{us}^2}{4\sin ^2\theta_{W}} \int\frac{d^4P_2}{(2\pi)^4}
\sum_{r_2,r_3} {\cal T}_{\mu\nu}^{r_3r_2}(\up_3,\up_2)
G_{+,r_3,s}^{>,<}(P_3)G_{+,r_2,u}^{<,>}(P_2).
\label{Pi-TwoTerms}
\end{equation}
where, by definition, the tensor ${\cal T}_{\mu\nu}^{rr^\prime}(\up,\up^\prime)$ is given by the 
following trace (in color and Dirac spaces):
\begin{equation}
{\cal T}_{\mu\nu}^{rr^\prime}(\up,\up^\prime)
={\rm Tr}[\gamma^{\mu}(1-\gamma^5)\gamma_0\Lambda _\mathbf{p}^-{\cal P}_{\mathbf{p},r}^+
          \gamma^{\nu}(1-\gamma^5)\gamma_0\Lambda _{\mathbf{p}^\prime}^-{\cal P}_{\mathbf{p}^\prime,r^\prime}^+]\, .
\label{definition-T}
\end{equation}
This trace was calculated for each of the four spin-one color superconducting phases in 
Ref.~\cite{Schmitt:2005wg}. For convenience, the corresponding results are also quoted 
in Appendix~\ref{app1}. 

Finally, by making use of Eqs.~(\ref{Ggtr}) and (\ref{Gless}), we arrive at the following expression
for the $W$-boson polarization tensor:
\begin{eqnarray}
\Pi_{\mu\nu}^{<,>} &=& \frac{i \pi e^2 V_{us}^2}{2\sin ^2\theta_{W}} \int\frac{d^3{\bf p}_2}{(2\pi)^3}
\sum_{r_2,r_3,e_1,e_2}{\cal T}_{\mu\nu}^{r_3r_2}(\up_3,\up_2)
B_{\bp_3,r_3,s}^{e_1} B_{\bp_2,r_2,u}^{e_2} \nonumber\\
&\times & f(\pm e_1 \epsilon_{\bp_3,r_3,s}) f(\mp e_2 \epsilon_{\bp_2,r_2,u})
\delta(q_0+\delta \mu -e_1 \epsilon_{\bp_3,r_3,s}+e_2\epsilon_{\bp_2,r_2,u}) .
\label{W-polarization}
\end{eqnarray}
Here we denote $\delta \mu\equiv \mu_s-\mu_u$ and assume that the upper (lower) sign corresponds to 
$\Pi^<$ ($\Pi^>$). It should be mentioned that one of the $\delta$-functions was used to perform the 
integration over $p_{2,0}$.

\section{Calculation of the rate}
\label{Calculation}

In this section we derive a general expression for the net rate of the $d$-quark production 
in spin-one color superconducting quark matter close to chemical equilibrium. The corresponding 
rate is formally defined by Eq.~(\ref{rate-Gamma}). By making use of the quark propagators and 
the $W$-boson polarization tensor, derived in the previous section, we obtain
\begin{eqnarray}
\Gamma_d &=& \frac{ie^2V_{ud}^2}{16 M_W^4\sin^2\theta_W}
\int\frac{d^4 P_1}{(2\pi)^4}\int\frac{d^4 P_4}{(2\pi)^4} 
\sum_{r_1,r_4} {\cal T}^{\mu\nu}_{r_4r_1}(\up_4,\up_1) \nonumber \\ 
&\times&  
\Big[  G_{+,r_1,d}^>(P_1)G_{+,r_4,u}^<(P_4) \Pi^>_{\mu\nu}(Q) 
    -  G_{+,r_4,u}^>(P_4)G_{+,r_1,d}^<(P_1) \Pi^<_{\mu\nu}(Q) \Big].
\label{rate-Gamma_2}
\end{eqnarray}
where we used the following results for the traces:
\begin{eqnarray}
{\rm Tr}\left( S_d^>(P_1)\Gamma_{ud,-}^\mu S_u^<(P_4)\Gamma_{ud,+}^\nu\right) &=& \frac{e^2V_{ud}^2}{4\sin^2\theta_W} 
\sum_{r_1,r_4} {\cal T}^{\mu\nu}_{r_4r_1}(\up_4,\up_1) G_{+,r_1,d}^>(P_1)G_{+,r_4,u}^<(P_4)  ,
\\
{\rm Tr}\left(\Gamma_{ud,-}^\mu S_u^>(P_4) \Gamma_{ud,+}^\nu S_d^<(P_1)\right) &=& \frac{e^2V_{ud}^2}{4\sin^2\theta_W}
\sum_{r_1,r_4} {\cal T}^{\mu\nu}_{r_4r_1}(\up_4,\up_1) G_{+,r_4,u}^>(P_4)G_{+,r_1,d}^<(P_1)  .
\end{eqnarray}
As in the calculation of the polarization tensor, the anomalous (off-diagonal) Nambu-Gorkov components of 
quark propagators did not contribute to these traces. This is the consequence of the specific flavor 
structure of the weak interaction vertices (\ref{Gamma-ud-us}).

After making use of Eqs.~(\ref{Ggtr}), (\ref{Gless}) and (\ref{W-polarization}), we obtain
\begin{eqnarray}
\Gamma_d &=& 2^7 \pi^4 G_F^2V^2_{ud}V^2_{us}\sum_{r_1 r_2 r_3 r_4}\sum_{e_1 e_2 e_3 e_4}
\int\frac{d^3{\bf p}_1}{(2\pi)^3}\frac{d^3{\bf p}_2}{(2\pi)^3}\frac{d^3{\bf p}_3}{(2\pi)^3}
\frac{d^3{\bf p}_4}{(2\pi)^3}(1-\up_1\cdot\up_2)(1-\up_3\cdot\up_4)
\omega_{r_4r_1}(\up_4,\up_1) \omega_{r_3r_2}(\up_3,\up_2)
\nonumber \\ 
& \times &
B_{\bp_1,r_1,d}^{e_1} B_{\bp_2,r_2,u}^{e_2} B_{\bp_3,r_3,s}^{e_3} B_{\bp_4,r_4,u}^{e_4} 
\delta({\bf p}_1+{\bf p}_2-{\bf p}_3-{\bf p}_4) 
\delta(e_1\epsilon_{\bp_1,r_1,d}+e_2\epsilon_{\bp_2,r_2,u}-e_3\epsilon_{\bp_3,r_3,s}-e_4\epsilon_{\bp_4,r_4,u}+ \delta \mu) \nonumber \\ 
& \times &
\left[f( e_1\epsilon_{\bp_1,r_1,d}) f( e_2\epsilon_{\bp_2,r_2,u}) f(-e_3\epsilon_{\bp_3,r_3,s}) f(-e_4\epsilon_{\bp_4,r_4,u})
    - f(-e_1\epsilon_{\bp_1,r_1,d}) f(-e_2\epsilon_{\bp_2,r_2,u}) f( e_3\epsilon_{\bp_3,r_3,s}) f( e_4\epsilon_{\bp_4,r_4,u})\right] .
\nonumber \\ 
\label{inte}
\end{eqnarray}
\end{widetext}
In derivation, we used the definition of the Fermi constant in terms of the $W$-boson mass,
\begin{equation}
G_F=\frac{e^2}{4\sqrt{2}\sin^2\theta_W M_W^2} \,  
\end{equation}
and the following Lorentz contraction:
\begin{eqnarray}
{\cal T}^{\mu\nu}_{r_4r_1}(\up_4,\up_1) {\cal T}^{r_3r_2}_{\mu\nu}(\up_3,\up_2)
=16(1-\up_1\cdot\up_2) \qquad\quad \nonumber\\
\times (1-\up_3\cdot\up_4) \omega_{r_4r_1}(\up_4,\up_1) \omega_{r_3r_2}(\up_3,\up_2)\, ,
\end{eqnarray}
where $\omega_{rr^\prime}(\up,\up^\prime)$ denotes a color trace that involves a pair of quasiparticles 
($r$ and $r^\prime$) with the given directions of their three-momenta ($\up$ and $\up^\prime$) in a 
specific spin-one color superconducting phase. The corresponding traces for all four phases 
are listed in Appendix~\ref{app1}. 

Formally, the expression in Eq.~(\ref{inte}) gives the net rate of the $d$-quark production 
in quark matter away from chemical equilibrium. The first term in the brackets 
describes the production of $d$-quarks due to $s+u\rightarrow u+d$, while the second one 
describes the annihilation of $d$-quarks due to $u+d\rightarrow s+u$. 

Here it might be instructive to note that the above expression for the rate $\Gamma_d$ resembles
the general result for the net rate of the $d$-quark production in the normal phase of strange 
quark matter \cite{Madsen:1993xx}. The key difference comes from the presence of the Bogoliubov 
coefficients $B_{\bp,r,f}$ and the $\omega_{rr^\prime}(\up,\up^\prime)$ functions that account 
for a non-trivial quark structure of the quasiparticles in spin-one color superconductors. 
Naturally, when such quasiparticles are the asymptotic states for the weak processes, the 
amplitude is not the same as in the normal phase.

The degree of departure from $\beta$-equilibrium and, thus, the net rate is controlled by the 
parameter $\delta\mu=\mu_s-\mu_d$. When $\delta\mu=0$, the expression in the square brackets 
of Eq.~(\ref{inte}) vanishes and $\Gamma_d=0$. When $\delta\mu\neq 0$, on the other hand, 
one has
\begin{equation}
\Gamma_d\simeq \lambda \delta\mu
\end{equation}
to leading order in small $\delta\mu$ \cite{footnote}. Note that the overall sign was chosen so that 
$\lambda$ is positive definite. (Recall that a positive $\delta\mu$ means an excess of strange quarks, 
which should drive a net production of $d$-quarks, while a negative $\delta\mu$ means a deficit
of strange quarks, which will be produced by annihilating some $d$-quarks.)

From the general expression in Eq.~(\ref{inte}), we derive 
\begin{widetext}
\begin{eqnarray}
\lambda &=& \frac{5\lambda_0}{2^{11} \pi^5 \mu^5 T^3}\sum_{r_1 r_2 r_3 r_4}\sum_{e_1 e_2 e_3 e_4}
\int d^3{\bf p}_1 d^3{\bf p}_2 d^3{\bf p}_3 d^3{\bf p}_4 (1-\up_1\cdot\up_2) (1-\up_3\cdot\up_4)
\omega_{r_4r_1}(\up_4,\up_1) \omega_{r_3r_2}(\up_3,\up_2)
\nonumber \\ 
& \times &
B_{\bp_1,r_1,d}^{e_1} B_{\bp_2,r_2,u}^{e_2} B_{\bp_3,r_3,s}^{e_3} B_{\bp_4,r_4,u}^{e_4} 
\delta({\bf p}_1+{\bf p}_2-{\bf p}_3-{\bf p}_4) 
\delta(e_1\epsilon_{\bp_1,r_1,d}+e_2\epsilon_{\bp_2,r_2,u}-e_3\epsilon_{\bp_3,r_3,s}-e_4\epsilon_{\bp_4,r_4,u}) \nonumber \\ 
& \times &
f(-e_1\epsilon_{\bp_1,r_1,d}) f(-e_2\epsilon_{\bp_2,r_2,u}) f( e_3\epsilon_{\bp_3,r_3,s}) f( e_4\epsilon_{\bp_4,r_4,u}) .
\label{lambda-1}
\end{eqnarray}
where 
\begin{equation}
\lambda_0 = \frac{64G_F^2V^2_{ud}V^2_{us}}{5\pi^3}\mu^5 T^2
\label{lambda0}
\end{equation}
is the corresponding $\lambda$-rate in the normal phase of strange quark matter \cite{Madsen:1993xx}.

\subsection{Analysis of the rate in CSL phase}
\label{CSLPhase}

Out of the four spin-one color superconducting phases studied in this paper, the CSL phase is special. 
This is the only phase in which the dispersion relations of quasiparticles are isotropic. As a result, 
the corresponding rate is the easiest to calculate. In this subsection, we analyze the $\lambda$-rate 
in the CSL phase in detail. 

Let us start by noting that the explicit form of the $\omega_{rr^\prime}(\up,\up^\prime)$-functions 
in the CSL phase is given by
\begin{eqnarray}
\omega_{11}(\up,\up^\prime) = 1+\frac{1}{4}(1+\up\cdot\up^\prime)^2\, ,
\label{ome11}\\
\omega_{12}(\up,\up^\prime) = \omega_{21}(\up,\up^\prime)
                            =1-\frac{1}{4}(1+\up\cdot\up^\prime)^2\, ,
\label{ome12}\\
\omega_{22}(\up,\up^\prime) = \frac{1}{4}(1+\up\cdot\up^\prime)^2\, . 
\label{ome22}
\end{eqnarray}
(See Appendix~\ref{app1} and Ref.~\cite{Schmitt:2005wg}.)
By making use of these expressions and introducing the following notation for the angular integrals:
\begin{equation}
F_{r_1 r_2 r_3 r_4} = \int d\Omega_1\int d\Omega_2 \int d\Omega_3 \int d\Omega_4\, 
(1-{\hat {\bf p}}_3\cdot{\hat {\bf p}}_4)  
(1-{\hat {\bf p}}_1\cdot{\hat {\bf p}}_2) 
\omega_{r_4r_1}(\up_4,\up_1) \omega_{r_3r_2}(\up_3,\up_2) 
\delta({\bf p}_1+{\bf p}_2-{\bf p}_3-{\bf p}_4) 
\, , \label{fs}
\end{equation}
we arrive at the following representation for the $\lambda$-rate in the CSL phase:
\begin{eqnarray}
\hspace*{-0.12in}
\lambda^{\rm (CSL)} &=& \frac{5\lambda_0\mu^3 }{2^{11} \pi^5 T^3}\sum_{r_1 r_2 r_3 r_4}\sum_{e_1 e_2 e_3 e_4}
\int_{0}^{\infty} d{p}_1 \int_{0}^{\infty} d{p}_2 \int_{0}^{\infty} d{p}_3 \int_{0}^{\infty} d{p}_4  \, 
F_{r_1 r_2 r_3 r_4} B_{\bp_1,r_1,d}^{e_1} B_{\bp_2,r_2,u}^{e_2} B_{\bp_3,r_3,s}^{e_3} B_{\bp_4,r_4,u}^{e_4}
\nonumber \\ 
& \times &
f(-e_1\epsilon_{\bp_1,r_1,d}) f(-e_2\epsilon_{\bp_2,r_2,u}) f( e_3\epsilon_{\bp_3,r_3,s}) f( e_4\epsilon_{\bp_4,r_4,u}) 
\delta(e_1\epsilon_{\bp_1,r_1,d}+e_2\epsilon_{\bp_2,r_2,u}-e_3\epsilon_{\bp_3,r_3,s}-e_4\epsilon_{\bp_4,r_4,u}) .
\label{lambdaCSL}
\end{eqnarray}
\end{widetext}
Here we took into account that, to leading order in inverse powers of $\mu$, the absolute values of 
the quark three-momenta can be approximated by $\mu$. In the same approximation, the explicit form
of functions $F_{r_1 r_2 r_3 r_4}$ are given in Appendix~\ref{app2}. All of them are proportional 
to $1/\mu^3$. This factor cancels out with the overall $\mu^3$ in Eq.~(\ref{lambdaCSL}). In order 
to perform the remaining numerical integrations, it is convenient to introduce new dimensionless 
integration variables $x_i=(p_i-\mu)/T$ instead of $p_i$ ($i=1,2,3,4$). The integration over $x_4$ 
is done explicitly by making use of the $\delta$-function. The remaining three-dimensional 
integration is done numerically, using a Monte-Carlo method. One finds that the ratio 
$\lambda^{\rm (CSL)}/\lambda_{0}$ is a function of a single dimensionless ratio, $\phi/T$.

Before proceeding to the numerical results, it is instructive to analyze the limiting case 
of low temperatures (or alternatively very large $\phi/T$). In this limit, only the ungapped 
$r=2$ quasiparticle modes should contribute to the rate. The corresponding contribution is easy 
to obtain analytically, i.e.,
\begin{eqnarray}
\lambda^{\rm (CSL)} &\simeq &  \frac{\lambda_0  F_{2222}}{\sum_{r_1 r_2 r_3 r_4} F_{r_1 r_2 r_3 r_4}} 
\nonumber\\
&=& \frac{928}{27027} \lambda_0
\approx 0.034 \lambda_0,
\quad \mbox{for} \quad
\frac{\phi}{T}\to\infty .
\label{rate2222}
\end{eqnarray}
The subleading correction to this result is suppressed by an exponentially small factor 
$\exp(-\sqrt{2}\phi/T)$. (Note that $\sqrt{2}$ in the exponent is connected with the  
conventional choice of the CSL gap, which is $\sqrt{2}\phi$ rather than $\phi$.) 

It might be instructive to mention that the asymptotic value in Eq.~(\ref{rate2222}) 
is substantially smaller than $\lambda_0/9$, which is the corresponding contribution 
of a single ungapped mode in the normal phase. The additional suppression comes from 
the functions $\omega_{22}(\up_4,\up_1)$ and $\omega_{22}(\up_3,\up_2)$ which modify 
the amplitude of the weak processes with respect to the normal phase. Except for the 
special case of collinear processes (i.e., $\up_4$ parallel to $\up_1$ and $\up_3$ 
parallel to $\up_2$), the corresponding $\omega$-functions are less than $1$, see 
Eq.~(\ref{ome22}). Interestingly, this kind of suppression is a unique property of 
the non-leptonic rates and is not seen in analogous Urca rates because the latter 
are dominated by the collinear processes \cite{Schmitt:2005wg,Sa'd:2006qv}. 

\begin{figure}[t]
\includegraphics[width=0.45\textwidth]{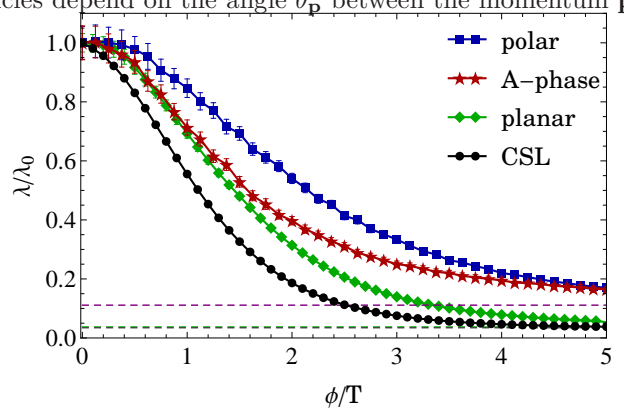}
\caption{(Color online) Numerical results for the $\lambda$-rate in four different phases 
of spin-one color superconducting strange quark matter \cite{EPAPS}. The error bars show 
the statistical error estimates in the Monte-Carlo calculation of the rates. The horizontal 
dashed lines correspond to the contributions of the ungapped modes in the limit of large 
$\phi/T$ (or equivalently the limit of low temperatures).}
\label{figAllPhases}
\end{figure}

All our numerical results for the $\lambda$-rates as a function of $\phi/T$ are shown 
in Fig.~\ref{figAllPhases} \cite{EPAPS}. In the case of the CSL phase (black points and 
the interpolating line in Fig.~\ref{figAllPhases}), we used the Mathematica's adaptive 
quasi-Monte-Carlo method to calculate the $\lambda$-rate. In order to improve the 
efficiency of the method, we partitioned the range of integration for each of the three 
dimensionless integration variables $x_i=(p_i-\mu)/T$ into several (up to 6) non-overlapping 
regions. This approach insures that the main contribution, coming from a close neighborhood 
of the Fermi sphere, is not lost in the integration over a formally very large phase space.

As seen from Fig.~\ref{figAllPhases}, the numerical results smoothly interpolate between the value 
of the rate in the normal phase $\lambda_0$ and the asymptotic value of the rate due to the CSL 
ungapped modes, given by Eq.~(\ref{rate2222}).

\subsection{Analysis of the rate in polar phase}
\label{PolarPhase}

Unlike the CSL phase, the polar phase is not isotropic. However, it is the simplest one among the 
other three phases. While the dispersions relations of its quasiparticles depend on the angle 
$\theta_{\bf p}$ between the momentum ${\bf p}$ and a fixed $z$-drection, its 
$\omega_{rr^\prime}(\up,\up^\prime)$-functions are independent of the quasiparticle momenta, i.e.,
\begin{equation}
\omega_{rr^\prime}(\up,\up^\prime) = n_{r}\delta_{rr^\prime} \,,
\end{equation}
with $n_1=2$ and $n_2=1$, see Appendix~\ref{app1}. Taking this into account, the corresponding 
$\lambda$-rate takes a simple form:
\begin{widetext}
\begin{eqnarray}
\lambda^{\rm (polar)} &=& \frac{5\lambda_0}{2^{11} \pi^5 \mu^5 T^3}\sum_{r_1 r_2}n_{r_1}n_{r_2}\sum_{e_1 e_2 e_3 e_4}
\int d^3{\bf p}_1 d^3{\bf p}_2 d^3{\bf p}_3 d^3{\bf p}_4 (1-\up_1\cdot\up_2) (1-\up_3\cdot\up_4)
\delta({\bf p}_1+{\bf p}_2-{\bf p}_3-{\bf p}_4)
\nonumber \\ 
& \times &
B_{\bp_1,r_1,d}^{e_1} B_{\bp_2,r_2,u}^{e_2} B_{\bp_3,r_2,s}^{e_3} B_{\bp_4,r_1,u}^{e_4} 
f(-e_1\epsilon_{\bp_1,r_1,d}) f(-e_2\epsilon_{\bp_2,r_2,u}) f( e_3\epsilon_{\bp_3,r_2,s}) f( e_4\epsilon_{\bp_4,r_1,u})
\nonumber \\ 
& \times &
\delta(e_1\epsilon_{\bp_1,r_1,d}+e_2\epsilon_{\bp_2,r_2,u}-e_3\epsilon_{\bp_3,r_2,s}-e_4\epsilon_{\bp_4,r_1,u})  .
\label{lambdaPolar}
\end{eqnarray}
By making use of the first $\delta$-function, we easily perform the integration over $\bp_4$. We can also 
perform the integration over one of the remaining polar coordinates. This is possible because the integrand depends 
only on the two independent combinations of the polar angles, i.e., $\tilde\varphi_1=\varphi_1-\varphi_3$ and 
$\tilde\varphi_2=\varphi_2-\varphi_3$. By using $\tilde\varphi_1$ and $\tilde\varphi_2$ as new integration 
variables (for simplicity of notation, the tildes are dropped in the following), we see that the integrand 
is independent of the variable $\varphi_3$. Finally, by approximating $p_1^2p_2^2p_3^2\simeq \mu^6$ in the 
integration measure, we rewrite the expression for the rate as follows:
\begin{eqnarray}
\lambda^{\rm (polar)} &\simeq & \frac{5\lambda_0 \mu}{2^{18} \pi^4 T}\sum_{r_1 r_2}n_{r_1}n_{r_2}
\int_{-\infty}^{\infty} d x_1 \int_{-\infty}^{\infty} d x_2 \int_{-\infty}^{\infty} d x_3 
\int_{-1}^{1} d \xi_1 \int_{-1}^{1} d \xi_2 \int_{-1}^{1} d \xi_3 
\int_{0}^{2\pi} d \varphi_1 \int_{0}^{2\pi} d \varphi_2 \nonumber \\ 
&\times& \frac{(1-\cos\theta_{12}) (1-\cos\theta_{34})}
{\cosh(\frac12 \epsilon_{x_1,\xi_1,r_1} )
 \cosh(\frac12 \epsilon_{x_2,\xi_2,r_2} )
 \cosh(\frac12 \epsilon_{x_3,\xi_3,r_2} )
 \cosh(\frac12 \epsilon_{x_4,\xi_4,r_1} ) }
\sum_{e_1 e_2 e_3 e_4}
 B_{x_1,\xi_1,r_1}^{e_1}
 B_{x_2,\xi_2,r_2}^{e_2}
 B_{x_3,\xi_3,r_2}^{e_3}
 B_{x_4,\xi_4,r_1}^{e_4}
\nonumber \\ 
& \times &
\delta(e_1\epsilon_{x_1,\xi_1,r_1}+e_2\epsilon_{x_2,\xi_2,r_2}-e_3\epsilon_{x_3,\xi_3,r_2}-e_4\epsilon_{x_4,\xi_4,r_1}) ,
\label{lambdaPolar2}
\end{eqnarray}
\end{widetext}
where the new integration variables are $x_i=(p_i-\mu)/T$ and $\xi_i = \cos\theta_{{\bf p}_i}$. 
By definition, the dimensionless energy is 
\begin{equation}
\epsilon_{x,\xi,r} = \sqrt{x^2+|\phi/T|^2 \lambda_{\xi,r}},
\end{equation} 
with $\lambda_{\xi,1}=1-\xi^2$ and $\lambda_{\xi,2}=0$, and the new Bogoliubov coefficients are
\begin{equation}
B_{x,\xi,r}^{e} = \frac{1}{2}\left(1-e\frac{x}{\epsilon_{x,\xi,r}}\right).
\end{equation} 
Note that the expressions for $x_4$, $\xi_4$ and $\cos\theta_{34}$ in Eq.~(\ref{lambdaPolar2}) are given by 
\begin{eqnarray}
x_4 &=&  \frac{p_4-\mu}{T}, \nonumber \\ 
\xi_4 &=&  \frac{p_1\xi_1+p_2\xi_2-p_3\xi_3}{p_4},\nonumber \\ 
\cos\theta_{34} &=& \frac{p_1\cos\theta_{13}+p_2\cos\theta_{23}-p_3}{p_4},
\end{eqnarray}
where $p_4=|\bp_1+\bp_2-\bp_3|$ is a function of $x_i$ ($i=1,2,3$) and the three cosine functions,
\begin{eqnarray}
\cos\theta_{12} &=& \xi_1\xi_2+\sqrt{1-\xi_1^2}\sqrt{1-\xi_2^2}\cos(\varphi_1-\varphi_2) , \nonumber \\ 
\cos\theta_{13} &=& \xi_1\xi_3+\sqrt{1-\xi_1^2}\sqrt{1-\xi_3^2}\cos(\varphi_1), \nonumber \\ 
\cos\theta_{23} &=& \xi_2\xi_3+\sqrt{1-\xi_2^2}\sqrt{1-\xi_3^2}\cos(\varphi_2) .
\end{eqnarray}
In the calculation, we used a customized Monte-Carlo method in order to improve the statistical 
error of the integration over $x_i$ (with $i=1,2,3$). To this end, we used a special type of 
importance sampling, which is motivated by the fact that the main contribution to the rate 
should come from the region near the Fermi surface. In order to implement this, we utilized 
random variables distributed according to the Gaussian distribution \cite{Weinzierl:2000wd}:
\begin{equation}
P(x)=\frac{1}{\sqrt{2\pi\sigma^2}}\exp\left(-\frac{(x-x_0)^2}{2\sigma^2}\right),
\label{GaussianDistribution}
\end{equation}
where $x_0$ and $\sigma$ are the mean and the width of the distribution, respectively. This was 
applied to the numerical integration over the dimensionless variables $x_i=(p_i-\mu)/T$ ($i=1,2,3$), 
in which case we took $x_0=0$ and $\sigma=3$. In order to generate independent variables (e.g., 
$x_1$ and $x_2$), distributed according to Eq.~(\ref{GaussianDistribution}), we applied the 
Box-Muller transform, 
\begin{eqnarray}
x_1 &=& x_0+\sigma \sqrt{-2\ln u_1} \cos (2\pi u_2),\\ 
x_2 &=& x_0+\sigma \sqrt{-2\ln u_1} \sin (2\pi u_2),
\end{eqnarray}
where $u_1$ and $u_2$ are two independent variables, uniformly distributed in the range from $0$ to $1$. 

In our numerical calculation, we also used a Gaussian function to approximate the $\delta$-function 
responsible for the energy conservation in the expression for the rate (\ref{lambdaPolar2}). For this 
purpose, we used the width of the distribution $\sigma_0=0.2$. This appeared to be sufficiently small 
to avoid strong violations of the energy conservation in the weak processes and, at the same time, 
sufficiently large to use in a Monte-Carlo integration with the number of (eight-dimensional) random 
points on the order of $10^6$ (in a Mathematica code) or $10^7$ (in a Fortran/C++ code).  

The numerical results for the $\lambda$-rate in the polar phase are shown by the blue squares (and 
the interpolating line) in Fig.~\ref{figAllPhases}. At vanishing $\phi/T$, the rate coincides with 
that in the normal phase. At asymptotically large value of $\phi/T$, on the other hand, the rate 
approaches $\lambda_0/9$. This value is marked by the purple dashed line in the figure. Theoretically,
the rate is dominated by the ungapped modes ($r_1=r_2=2$) in the $\phi/T\to \infty$ limit. The 
corresponding contribution can be obtained by analytical methods as follows. We start by pointing 
that the Bogoliubov coefficients for the ungapped modes are equal to the unit step functions: 
$B_{x_i,\xi_i,2}^{e_i} \equiv \Theta(-e_i x_i)$, where by definition $\Theta(x)=1$ for $x\geq 0$ and 
$\Theta(x)=0$ otherwise. Since these Bogoliubov coefficients are nonzero only for $e_i=\mbox{sign}(-x_i)$,
each sums over $e_i$ effectively reduces to a single contribution. By taking this into account
and making use of the result for the angular integration, $K_0$, defined in Appendix~\ref{app2}, 
we derive  
\begin{eqnarray}
\lambda^{\rm (polar)}_{\rm unpaired} &\simeq & \frac{\lambda_0}{6 \pi^2} 
\int_{-\infty}^{\infty} d x_1 \int_{-\infty}^{\infty} d x_2 
\int_{-\infty}^{\infty} d x_3 \int_{-\infty}^{\infty} d x_4 \nonumber\\
&\times& \frac{\delta(-x_1-x_2+x_3+x_4)}
{(e^{x_1}+1)(e^{x_2}+1)(e^{-x_3}+1)(e^{-x_4}+1)} \nonumber\\
&=& \frac{1}{9} \lambda_0.
\label{lambdaPolar3}
\end{eqnarray}
It should be noted that the numerical results for the polar phase in Fig.~\ref{figAllPhases} 
approach this asymptotic value very slowly. We can speculate that this indicates a weak 
(probably, power-law) suppression of the contribution of the gapped (mixed with ungapped) modes  
to the rate. The key feature responsible for this behavior in the polar phase is the presence 
of gapless nodes at $\theta_{\bf p} =0$ and $\theta_{\bf p} =\pi$ in the dispersion relation 
of the gapped modes. As we shall see below, the same qualitative property is shared by the 
{\em A}-phase, whose gapped modes also have a node at $\theta_{\bf p} =\pi$. In contrast, the 
rates in the CSL and planar phases, whose gapped modes have no gapless nodes, show 
asymptotes that are consistent with the rapid, exponential approach to their asymptotic 
values.

\subsection{Analysis of the rate in {\em A}-phase}
\label{Aphase}

The analysis in the {\em A}-phase of spin-one color superconducting matter can be performed along the
same lines as in the polar phase. The apparent complication of the {\em A}-phase is the existence of 
three, rather than two distinct quasiparticle excitations. However, it appears that the contributions 
of the two gapped modes ($r=1,2$) can be replaced by a single contribution of a modified mode with the 
energy $\epsilon_{\bf p}=\sqrt{(p-\mu)^2+|\phi|^2\lambda_{\bf p}}$ where $\lambda_{\bf p}\equiv 
(1+\cos\theta_{\bf p})^2$ (cf. the dispersion relations of the modes $r=1,2$ in Tab.~\ref{tablephases}). 
This alternative representation is possible because of the special, separable structure of the corresponding 
$\omega_{rr^\prime}(\up,\up^\prime)$ functions in the {\em A}-phase. As seen from the expressions in 
Eq.~(\ref{omegaAphase}), the mode $r=1$ contributes only when $\cos\theta_{\bf p}$ of the corresponding 
quasiparticle is positive, while the mode $r=2$ contributes only when $\cos\theta_{\bf p}$ is negative. 
Then, when the contributions are nonvanishing, one always gets $\omega_{rr^\prime}(\up,\up^\prime)=2$. 
By also noting that the corresponding eigenvalues 
\begin{equation}
\lambda_{{\bf p},1} =(1+|\cos\theta_{\bf p}|)^2\quad \mbox{for}\quad\cos\theta_{\bf p}>0
\end{equation}
and
\begin{equation}
\lambda_{{\bf p},2} =(1-|\cos\theta_{\bf p}|)^2\quad \mbox{for}\quad\cos\theta_{\bf p}<0
\end{equation}
formally take the same form, i.e., $\lambda_{\bf p}\equiv (1+\cos\theta_{\bf p})^2$, we conclude
that the sum over the original modes $r=1,2$ in the rate can indeed be replaced by a single contribution
of the modified mode as defined above. By making use of this observation, the general expression for 
the rate in the {\em A}-phase takes the form, which is similar to that in the polar phase, see 
Eq.~(\ref{lambdaPolar}), but with a different dispersion relation of the (modified) gapped mode.   

By using a Monte-Carlo algorithm as in the previous case, we perform a numerical calculation of 
the $\lambda$-rate in the {\em A}-phase. The corresponding results are shown by red stars (and 
the interpolating line) in Fig.~\ref{figAllPhases}. In the limit of large $\phi/T$, the rate is 
saturated by the contribution of ungapped modes, which is the same as in the polar phase, namely 
$\lambda_0/9$. The derivation of this asymptotic expression is the same as in the polar phase.
The corresponding value is marked by the purple dashed line in the figure. A slow (probably, 
power-law) approach of the asymptotic value at $\phi/T\to \infty$ is again associated with the
presence of a gapless node (at $\theta_{\bf p} =\pi$) in the dispersion relation of the (modified) 
gapped quasiparticles.

\subsection{Analysis of the rate in planar phase}
\label{PlanarPhase}

The calculation of the rate in the planar case requires the largest amount of computer 
time. One of the main reasons for that is the much more complicated expressions for the 
$\omega_{rr^\prime}(\up,\up^\prime)$-functions (see Appendix~\ref{app1}). The numerical 
results for the $\lambda$-rate in the planar phase are shown by green diamonds (and the 
interpolating line) in Fig.~\ref{figAllPhases}. The asymptotic
value of the rate at large $\phi/T$ was extracted only numerically. By taking into account 
possible systematic errors (e.g., due to the overall normalization of the rate that may differ
by up to 15\% from the analytical estimate (\ref{lambda0}) in the normal phase), we estimate 
$\lambda^{\rm (planar)}\simeq (0.038\pm 0.003)\lambda_0$ for $\phi/T\to\infty$. Note that this
is smaller than $\lambda_0/9$, which is the contribution of a single mode in the normal phase. 
As in the CSL phase,  in the planar phase the additional suppression comes from the 
$\omega$-functions for the ungapped modes.

\section{Discussion}
\label{Discussion}

In this paper we derived the near-equilibrium rates of the net $d$-quark production 
(or equivalently the $\lambda$-rates) due to the non-leptonic weak processes (i.e., 
the difference of the rates of $u + d \rightarrow u + s$ and $u + s \rightarrow u + d$) 
in spin-one color-superconducting strange quark matter at high density. The main 
numerical results are presented in Fig.~\ref{figAllPhases}.

In the limit of $\phi/T=0$, which is same as the normal (unpaired) phase of strange quark 
matter, our results reproduce the known result of Ref.~\cite{Madsen:1993xx}. The effect of 
color superconductivity is to suppress these rates. The degree of the suppression depends 
on the details of the specific spin-one phases. To large extent, this is controlled by the 
value of the energy gap (more precisely, $\phi/T$) as well as its functional dependence 
on the direction of the quasiparticle momentum. At very large $\phi/T$ (or equivalently 
in the limit of low temperatures), the $\lambda$-rates approach fixed values, which are 
determined by the contribution of the ungapped modes alone. The corresponding limiting 
value is the smallest in the CSL phase. It is less than a third of the ``canonical" 
value $\lambda_0/9$ due to a single ungapped mode in the normal phase of matter. The
additional suppression comes from the modification of the quasiparticles due to color
superconductivity. A similar observation applies to the planar phase. The rates in the 
other two phases, i.e., the polar and {\em A}-phase, approach the asymptotic
values equal to $\lambda_0/9$.

The numerical results for the $\lambda$-rates in Fig.~\ref{figAllPhases} also indicate 
that the asymptotic approach to the limiting values can be qualitatively different in 
spin-one color superconducting phases. In the case of the polar and {\em A}-phase, the 
approach seems to follow a power law. In contrast, the approach appears to be exponential 
in the case of the CSL and planar phase. This qualitative difference can be easily 
understood. The power law is the consequence of the presence of gapless nodes in the 
dispersion relations of the gapped quasiparticles in the polar and {\em A}-phase (the 
nodes are located at $\theta_{\bf p}=0$ and $\theta_{\bf p}=\pi$ in the polar phase,
and at $\theta_{\bf p}=\pi$ in the {\it A}-phase). In the CSL and planar phase, the 
approach to the asymptotic value at $\phi/T\to\infty$ is exponential because no gapless 
nodes are found in their gapped quasiparticles. (Note that a similar observation regarding 
the rates of the semi-leptonic processes was made in Ref.~\cite{Schmitt:2005wg,Sa'd:2006qv}.)

The results for the rates of non-leptonic weak processes, presented here, is an important 
ingredient for the calculation of the bulk viscosity of spin-one color-superconducting strange 
quark matter. If such matter is present inside neutron stars, its viscosity will be one of 
the mechanisms responsible for damping of the stellar $r$-mode instabilities \cite{Andersson:1997xt}.

\acknowledgments 
The authors would like thank Mark Alford, Prashanth Jaikumar, Armen Sedrakian, Andreas Schmitt 
and Qun Wang for useful comments. 
H.M. acknowledges discussions with Lars Zeidlewicz. The work of I.A.S. was supported in part 
by the start-up funds from the Arizona State University.

\appendix

\section{Color and Dirac traces}
\label{app1}

In this appendix, we write down the explicit expressions for the tensor 
${\cal T}_{\mu\nu}^{rr^\prime}(\up,\up^\prime)$, defined by Eq.~(\ref{definition-T}). 
The corresponding results were obtained in Ref.~\cite{Schmitt:2005wg}. In 
general, one finds that  
\begin{equation}
{\cal T}^{\mu\nu}_{r,r^\prime}(\up,\up^\prime)
={\cal T}^{\mu\nu} (\up,\up^\prime) \omega_{rr^\prime}(\up,\up^\prime)\, ,
\label{Tmunu}
\end{equation}
where $\omega_{rr^\prime}(\up,\up^\prime)$ are the functions determined by a 
specific color-spin structure of the gap matrix, and 
\begin{equation}
{\cal T}^{\mu\nu} (\up,\up^\prime) \equiv \mbox{Tr}_D\left[
\gamma^\mu (1-\gamma^5)\gamma^0\Lambda_{\bf p}^- 
\gamma^\nu (1-\gamma^5)\gamma^0\Lambda_{\bf p^\prime}^-\right]. 
\label{T0munu}
\end{equation}
The explicit form of all the components of this tensor can be also found in 
Ref.~\cite{Schmitt:2005wg}. It is more important for us here to note that the 
following result for the contraction of this tensor with itself is valid:
\begin{equation}
{\cal T}^{\mu\nu} (\up_4,\up_1) {\cal T}_{\mu\nu}(\up_3,\up_2)
=16(1-\up_1\cdot\up_2)(1-\up_3\cdot\up_4) \, .
\end{equation}
Since an essential information regarding spin-one color-superconducting phases 
is carried by the $\omega_{rr^\prime}(\up,\up^\prime)$ functions, we also quote
them here. (For more details, see Ref.~\cite{Schmitt:2005wg}.)

In the {\em polar} phase, the $\omega_{rr^\prime}(\up,\up^\prime)$ functions do 
not depend on the quark momenta. They are given by the following expressions:
\begin{subequations}
\begin{eqnarray}
\omega_{11}(\up,\up^\prime) &=& 2 \,,\\
\omega_{22}(\up,\up^\prime) &=& 1 \,,\\
\omega_{12}(\up,\up^\prime) &=& \omega_{21}(\up,\up^\prime)=0 \,.
\end{eqnarray}
\end{subequations}

In the {\em planar} phase, the explicit form of the $\omega_{rr^\prime}(\up,\up^\prime)$ 
functions reads
\begin{subequations}
\begin{eqnarray} 
\omega_{11}(\up,\up^\prime)&=&\frac{1}{2}\,[3+\eta(\up,\up^\prime)]\,,\\
\omega_{12}(\up,\up^\prime)&=&\omega_{21}(\up,\up^\prime)=\frac{1}{2}\,[1-\eta(\up,\up^\prime)]\,,\\
\omega_{22}(\up,\up^\prime)&=&\frac{1}{2}\,[1+\eta(\up,\up^\prime)]\,, \label{ome22planar}
\end{eqnarray}
\end{subequations}
where
\begin{equation} \label{eta}
\eta(\up,\up^\prime)\equiv \frac{4\hp_z\hp_z^\prime + (\hp_x\hp_x^\prime+\hp_y\hp_y^\prime)^2 -(\hp_x\hp_y^\prime-\hp_y\hp_x^\prime)^2}
{[1+(\hp_z)^2][1+(\hp_z^\prime)^2]} \,\, .
\end{equation}

In the {\it A}-phase, there are three different quasiparticle branches ($r=1,2,3$). 
Consequently, there are more $\omega_{rr^\prime}(\up,\up^\prime)$ functions, i.e.,
\begin{subequations}
\begin{eqnarray}
&&
\omega_{11}(\up,\up^\prime)= \frac{1}{2}[1+{\rm sgn}(\hp_z)][1+{\rm sgn}(\hp_z^\prime)]  ,\\
&&
\omega_{22}(\up,\up^\prime)= \frac{1}{2}[1-{\rm sgn}(\hp_z)][1-{\rm sgn}(\hp_z^\prime)] , \\
&&
\omega_{12}(\up,\up^\prime)= \frac{1}{2}[1+{\rm sgn}(\hp_z)][1-{\rm sgn}(\hp_z^\prime)]  ,  \\
&&
\omega_{21}(\up,\up^\prime)= \frac{1}{2}[1-{\rm sgn}(\hp_z)][1+{\rm sgn}(\hp_z^\prime)]  ,  \\
&&
\omega_{13}(\up,\up^\prime)=\omega_{31}(\up,\up^\prime)=0,  \\
&& 
\omega_{23}(\up,\up^\prime)=\omega_{32}(\up,\up^\prime)=0  ,  \\
&&
\omega_{33}(\up,\up^\prime)= 1  .
\end{eqnarray}
\label{omegaAphase}
\end{subequations}
Finally, in the {\em CSL} phase, the corresponding functions are
\begin{eqnarray}
\omega_{11}(\up,\up^\prime) &=& 1 + \frac{1}{4}(1+\up\cdot\up^\prime)^2 ,\\
\omega_{12}(\up,\up^\prime) &=& \omega_{21}(\up,\up^\prime) 
= 1 - \frac{1}{4}(1+\up\cdot\up^\prime)^2 ,  \\
\omega_{22}(\up,\up^\prime) &=& \frac{1}{4}(1+\up\cdot\up^\prime)^2 .
\end{eqnarray}

\section{Angular integrations in CSL phase}
\label{app2}

In the calculation of the $\lambda$-rate in the CSL phase, there are four different
types of angular integrations over the phase space of quark momenta. Thus, the results
for the $F_{r_1 r_2 r_3 r_4}$ functions, formally defined by Eq.~(\ref{fs}) in the main 
text, have the following general structure:
\begin{subequations}
\begin{eqnarray}
F_{1111} &=& K_0 + K_1 + K_2 + K_3\,,\\
F_{1112} &=& K_0 - K_1 + K_2 - K_3\,,\\
F_{1121} &=& K_0 + K_1 - K_2 - K_3\,,\\
F_{1122} &=& K_0 - K_1 - K_2 + K_3\,,\\
F_{1211} &=& K_0 + K_1 - K_2 - K_3\,,\\
F_{1212} &=& K_0 - K_1 - K_2 + K_3\,,\\
F_{1221} &=& K_2 + K_3\,,\\
F_{1222} &=& K_2 - K_3\,, 
\end{eqnarray}
\begin{eqnarray}
F_{2111} &=& K_0 - K_1 + K_2 - K_3\,,\\
F_{2112} &=& K_1 + K_3\,,\\
F_{2121} &=& K_0 - K_1 - K_2 + K_3\,,\\
F_{2122} &=& K_1 - K_3\,,\\
F_{2211} &=& K_0 - K_1 - K_2 + K_3\,,\\
F_{2212} &=& K_1 - K_3\,,\\
F_{2221} &=& K_2 - K_3\,,\\
F_{2222} &=& K_3\,,
\end{eqnarray}
\end{subequations}
where the four types of angular integrals are given by:
\begin{widetext}
\begin{eqnarray}
\hspace*{-0.3in}
K_0 &=& \int{d\Omega_1}\int{d\Omega_2}\int{d\Omega_3}\int{d\Omega_4}(1-\up_1\cdot\up_2)
(1-\up_3\cdot\up_4) \delta({\bf p}_1+{\bf p}_2-{\bf p}_3-{\bf p}_4) ,\\
\hspace*{-0.3in}
K_1 &=& \frac{1}{4}\int{d\Omega_1}\int{d\Omega_2}\int{d\Omega_3}\int{d\Omega_4}(1-\up_1\cdot\up_2)
(1-\up_3\cdot\up_4)(1+\up_4\cdot\up_1)^2\delta({\bf p}_1+{\bf p}_2-{\bf p}_3-{\bf p}_4) , \\
\hspace*{-0.3in}
K_2 &=& \frac{1}{4}\int{d\Omega_1}\int{d\Omega_2}\int{d\Omega_3}\int{d\Omega_4}(1-\up_1\cdot\up_2)
(1-\up_3\cdot\up_4)(1+\up_3\cdot\up_2)^2\delta({\bf p}_1+{\bf p}_2-{\bf p}_3-{\bf p}_4) , \\
\hspace*{-0.3in}
K_3 &=& \frac{1}{16}\int{d\Omega_1}\int{d\Omega_2}\int{d\Omega_3}\int{d\Omega_4}(1-\up_1\cdot\up_2)
(1-\up_3\cdot\up_4) (1+\up_4\cdot\up_1)^2(1+\up_3\cdot\up_2)^2\delta({\bf p}_1+{\bf p}_2-{\bf p}_3-{\bf p}_4)\,.
\label{orders2}
\end{eqnarray}
The result for $K_0$ was obtained in Ref.~\cite{Alford:2006gy}. It reads
\begin{equation}
K_0 = \frac{4\pi^3}{p_1^2\,p_2^2\,p_3^2\,p_4^2}\,L_0(p_{12},P_{12},p_{34},P_{34}) \,,
\label{K0-L0}
\end{equation}
where $p_{ij}\equiv |p_i-p_j|$, $P_{ij}\equiv p_i+p_j$, and 
\begin{eqnarray}
L_0(a,b,c,d)&\equiv& \Theta(c-a)\Theta(d-b)\Theta(b-c)
J_0(c,b,b,d) +\Theta(a-c)\Theta(d-b)
J_0(a,b,b,d) \nonumber \\ 
&+&\Theta(a-c)\Theta(b-d)\Theta(d-a)
J_0(a,d,b,d) + \Theta(c-a)\Theta(b-d)
J_0(c,d,b,d)\,,
\end{eqnarray}
which is given in terms of 
\begin{eqnarray}
J_0(a,b,c,d)&\equiv&\int_a^b dP\,(c^2-P^2)(d^2-P^2)=c^2d^2(b-a) - \frac{1}{3}(c^2+d^2)(b^3-a^3)+\frac{1}{5} (b^5-a^5) \, .
\end{eqnarray}
To leading order in powers of large $\mu$, this result simplifies to
\begin{equation}
L_0(0,2\mu,0,2\mu) = \frac{2^8\mu^5}{15}.
\end{equation}
By making use of Eq.~(\ref{K0-L0}), therefore, we obtain
\begin{equation}
K_0 \simeq \frac{4\pi^3}{\mu^8}\,L_0(0,2\mu,0,2\mu)= \frac{2^{10}\pi^3}{15 \mu^3}.
\label{I0}
\end{equation}
Using the same approach, in the following subsections we calculate the results for 
$K_1$, $K_2$, and $K_3$.

\subsection{Calculation of $K_1$}

Here we calculate the angular integral $K_1$. Following the approach of Ref.~\cite{Alford:2006gy}, 
we obtain
\begin{eqnarray}
K_1&=&\frac{1}{4}
\int{d\Omega_1}\int{d\Omega_2}\int{d\Omega_3}\int{d\Omega_4}
(1-\up_1\cdot\up_2)(1-\up_3\cdot\up_4)
(1+\up_4\cdot\up_1)^2\delta({\bf p}_1+{\bf p}_2-{\bf p}_3-{\bf p}_4)\nonumber \\ 
&=&\frac{1}{4p_2^2}\int{d\Omega_1}\int{d\Omega_3}\int{d\Omega_4}
\left[1-\frac{1}{p_2}(\up_1\cdot{\bf P}-p_1)\right]
(1-\up_3\cdot\up_4)(1+\up_4\cdot\up_1)^2\delta(p_2-\vert\bf{P}-\bf{p_1}\vert)\nonumber \\ 
&=&\frac{1}{4p_2^2}\int{d\Omega_3}\int{d\Omega_4}\int_0^{2\pi}{d\phi_1}\int_0^{\pi}d\theta_1\sin\theta_1
\left[1-\frac{1}{p_2}(P\cos\theta_1-p_1)\right]
(1-\up_3\cdot\up_4)\delta(p_2-\vert\bf{P}-\bf{p_1}\vert)\nonumber \\ 
&&\times
\left[1+\cos\theta_1\cos\theta_4+\sin\theta_1\sin\theta_4\cos(\phi_1-\phi_4)\right]^2 ,
\end{eqnarray}
where $\bf{P}=\bf{p}_3+\bf{p}_4$ and $P=|\bf{P}|$. In order to integrate over $\theta_1$, we choose 
the coordinate system so that the $z$-axis is along the vector $\bf{P}$. After making use of the 
$\delta$-function, we easily integrate over $\theta_1$ and arrive at the following result:
\begin{eqnarray}
K_1&=&\int{d\Omega_3}\int{d\Omega_4} \frac{P^2_{12}-P^2}{8 P p_1^2 p_2^2}(1-\up_3\cdot\up_4)
\Theta(P_{12}-P)\Theta(P-p_{12})\nonumber \\ 
& & \times \int_0^{2\pi}d\phi_1 [1+\cos\theta_1^*\cos\theta_4+\sin\theta_1^*\sin\theta_4\cos(\phi_1-\phi_4)]^2\nonumber \\ 
&=&\pi\int{d\Omega_3}\int{d\Omega_4}\frac{P^2_{12}-P^2}{8 P p_1^2 p_2^2}\Theta(P_{12}-P)\Theta(P-p_{12})(1-\up_3\cdot\up_4)
\Big[2(1+\cos\theta^*_1 \cos\theta_4)^2+(\sin\theta^*_1\sin\theta_4)^2\Big]\,,
\label{K1-intermed}
\end{eqnarray}
where $\cos\theta_1^*=(P^2+p_1^2-p_2^2)/2Pp_1$ and $\cos\theta_4=(\up_4\cdot\uP)=(P^2+p_4^2-p_3^2)/2Pp_4$. 
It may be appropriate to emphasize that the result of the last integration was presented in a form that 
independent of a specific choice of the coordinate system. As can be easily checked, the integrand in 
Eq.~(\ref{K1-intermed}) depends only on the relative angle $\theta_{34}$ between the vectors $\up_3$ 
and $\up_4$ (or equivalently only on the variable $P$). Therefore, while integrating over $\Omega_4$, 
we could fix the orientation of $\up_3$ arbitrarily. It is convenient to choose $\up_3$ as the z-axis 
and perform the integration over $\Omega_4$. The result is independent of the angular coordinates in 
$\Omega_3$. Thus, the remaining integration over $\Omega_3$ gives an extra factor $4\pi$. In the end, 
we arrive at
\begin{eqnarray}
K_1&=& \frac{\pi^3}{2 p_1^2 p_2^2 p_3^2 p_4^2} \int_{p_{34}}^{P_{34}} g(P) d P \,,
\end{eqnarray}
where 
\begin{eqnarray}
g(P) &=& (P^2_{12}-P^2)(P^2_{34}-P^2)\Theta(P_{12}-P)\Theta(P-p_{12})
\Bigg[2\left(1+\frac{(P^2+p_1^2-p_2^2)(P^2+p_4^2-p_3^2)}{4P^2p_1 p_4}\right)^2\nonumber \\ 
&&+\left(1-\frac{(P^2+p_1^2-p_2^2)^2}{4P^2p_1^2}\right)\left(1-\frac{(P^2+p_4^2-p_3^2)^2}{4P^2p_4^2}\right)\Bigg]\,.
\end{eqnarray}
Note that we changed the integration variable from $\theta_{34}$ to $P=\sqrt{p_3^2+p_4^2+2p_3p_4 \cos\theta_{34}}$.

The final result for $K_1$ can be conveniently given in the same form as $K_0$ in the previous section, 
i.e.,
\begin{equation}
K_1 = \frac{\pi^3}{2 p_1^2 p_2^2 p_3^2 p_4^2}\,L_1(p_{12},P_{12},p_{34},P_{34}) \, ,
\end{equation}
where, by definition,
\begin{eqnarray}
L_1(a,b,c,d)&\equiv& \Theta(c-a)\Theta(d-b)\Theta(b-c)
J_1(c,b,b,d) +\Theta(a-c)\Theta(d-b)
J_1(a,b,b,d) \nonumber \\ 
&+&\Theta(a-c)\Theta(b-d)\Theta(d-a)
J_1(a,d,b,d) + \Theta(c-a)\Theta(b-d)
J_1(c,d,b,d)\,,
\end{eqnarray}
and
\begin{eqnarray}
J_1(a,b,c,d)&=&\int_a^b dP (c^2-P^2)(d^2-P^2)\Bigg[2\left(1+\frac{(P^2+p_1^2-p_2^2)(P^2+p_4^2-p_3^2)}{4P^2p_1 p_4}\right)^2\nonumber \\ 
&&+\left(1-\frac{(P^2+p_1^2-p_2^2)^2}{4P^2p_1^2}\right)\left(1-\frac{(P^2+p_4^2-p_3^2)^2}{4P^2p_4^2}\right)\Bigg]\,.
\end{eqnarray}
To leading order in powers of large $\mu$, this result reduces to
\begin{equation}
L_1(0,2\mu,0,2\mu) = \frac{2^{11}\mu^5}{35} .\\
\end{equation}
This, in turn, gives
\begin{equation}
K_1 \simeq \frac{\pi^3}{2\mu^8}\,L_1(0,2\mu,0,2\mu)= \frac{2^{10}\pi^3}{35 \mu^3}.
\label{I1}
\end{equation}

\subsection{Calculation of $K_2$}

As is easy to see, the expression for $K_2$ can be obtained from $K_1$ by the following 
exchange of variables: $p_1\leftrightarrow p_2$ and $p_3\leftrightarrow p_4$. 
Thus, the result reads
\begin{equation}
K_2 = \frac{\pi^3}{2 p_1^2 p_2^2 p_3^2 p_4^2}\,L_2(p_{12},P_{12},p_{34},P_{34}) \, ,
\end{equation}
where
\begin{eqnarray}
L_2(a,b,c,d)&\equiv& \Theta(c-a)\Theta(d-b)\Theta(b-c)
J_2(c,b,b,d) +\Theta(a-c)\Theta(d-b)
J_2(a,b,b,d) \nonumber \\ 
&+&\Theta(a-c)\Theta(b-d)\Theta(d-a)
J_2(a,d,b,d) + \Theta(c-a)\Theta(b-d)
J_2(c,d,b,d)\,,
\end{eqnarray}
and
\begin{eqnarray}
J_2(a,b,c,d)&=&\int_a^b dP (c^2-P^2)(d^2-P^2)\Bigg[2\left(1+\frac{(P^2+p_2^2-p_1^2)(P^2+p_3^2-p_4^2)}{4P^2p_2 p_3}\right)^2\nonumber \\ 
&&+\left(1-\frac{(P^2+p_2^2-p_1^2)^2}{4P^2p_2^2}\right)\left(1-\frac{(P^2+p_3^2-p_4^2)^2}{4P^2p_3^2}\right)\Bigg]\,.
\end{eqnarray}
We also find that $K_2$ is identical to $K_1$ to leading order in powers of large $\mu$, i.e., 
\begin{equation}
K_2 \simeq \frac{2^{10}\pi^3}{35 \mu^3}.
\label{I2}
\end{equation}

\subsection{Calculation of $K_3$}\label{app-d}

Now we calculate the angular integral $K_3$. We start by using the same approach as in the calculation of $K_1$,
\begin{eqnarray}
K_3&=&\frac{1}{16 p_2^2}\int{d\Omega_1}\int{d\Omega_3}\int{d\Omega_4}
\left[1-\frac{1}{p_2}(\up_1\cdot{\bf P}-p_1)\right]
(1-\up_3\cdot\up_4)(1+\up_4\cdot\up_1)^2\nonumber \\ 
&\times& \left[1+\frac{1}{p_2 p_3}\left({\bf p}_3\cdot{\bf P}-{\bf p}_1\cdot{\bf P}+{\bf p}_1\cdot{\bf p}_4\right)\right]^2
\delta(p_2-\vert\bf{P}-\bf{p_1}\vert)\,.
\end{eqnarray}
To calculate the integral over $\Omega_1$, we fix the coordinate system so that the z-axis coincides with the direction 
of ${\bf P}$. After integration, we obtain
\begin{eqnarray}
K_3&=&\int{d\Omega_3}\int{d\Omega_4} \frac{P_{12}^2-P^2}{32 Pp_1^2 p_2^2}\Theta(P_{12}-P)\Theta(P-p_{12})
(1-\up_3\cdot\up_4)\nonumber \\ 
&\times& \int_0^{2\pi}d\phi_1  [1+\cos\theta_1^*\cos\theta_4+\sin\theta_1^*\sin\theta_4\cos(\phi_1-\phi_4)]^2\nonumber \\ 
&& \times \left[1+\frac{p_2^2-p_1^2+p_3^2-p_4^2}{2p_2 p_3}
+\frac{p_1 p_4}{p_2 p_3}\left(\cos\theta_1^*\cos\theta_4+\sin\theta_1^*\sin\theta_4\cos(\phi_1-\phi_4)\right)\right]^2,
\end{eqnarray}
where $\cos\theta_1^*=(P^2+p_1^2-p_2^2)/2Pp_1$ and $\cos\theta_4=(\up_4\cdot\uP)=(P^2+p_4^2-p_3^2)/2Pp_4$. 
In the derivation, we also used the following relation:
\begin{equation}
{\bf p}_3\cdot{\bf P} = \frac{P^2+p_3^2-p_4^2}{2}\, .
\end{equation}
By performing the integration over $\phi_1$, we derive 
\begin{eqnarray}
K_3&=& \pi \int{d\Omega_3}\int{d\Omega_4} \frac{(P_{12}^2-P^2)(P_{34}^2-P^2)}{256 P p_1^2 p_2^2 p_3^3 p_4^3}
\Theta(P_{12}-P)\Theta(P-p_{12})\nonumber \\ 
&\times&\Bigg\{ \left(P_{14}^2-P_{23}^2\right)^2
\Bigg[2\left(1+\frac{(P^2+p_1^2-p_2^2)(P^2+p_4^2-p_3^2)}{4P^2p_1 p_4}\right)^2
+\left(1-\frac{(P^2+p_1^2-p_2^2)^2}{4P^2p_1^2}\right)\left(1-\frac{(P^2+p_4^2-p_3^2)^2}{4P^2p_4^2}\right)\Bigg]
\nonumber \\ 
&-&4p_1p_4\left(P_{14}^2-P_{23}^2\right)\left(1+\frac{(P^2+p_1^2-p_2^2)(P^2+p_4^2-p_3^2)}{4P^2p_1 p_4}\right)\nonumber \\ 
&& \times\Bigg[2\left(1+\frac{(P^2+p_1^2-p_2^2)(P^2+p_4^2-p_3^2)}{4P^2p_1 p_4}\right)^2
+3\left(1-\frac{(P^2+p_1^2-p_2^2)^2}{4P^2p_1^2}\right)\left(1-\frac{(P^2+p_4^2-p_3^2)^2}{4P^2p_4^2}\right)\Bigg]
\nonumber \\ 
&+&p_1^2 p_4^2
\Bigg[8\left(1+\frac{(P^2+p_1^2-p_2^2)(P^2+p_4^2-p_3^2)}{4P^2p_1 p_4}\right)^4
+3\left(1-\frac{(P^2+p_1^2-p_2^2)^2}{4P^2p_1^2}\right)^2\left(1-\frac{(P^2+p_4^2-p_3^2)^2}{4P^2p_4^2}\right)^2\nonumber \\ 
&&+24 \left(1+\frac{(P^2+p_1^2-p_2^2)(P^2+p_4^2-p_3^2)}{4P^2p_1 p_4}\right)^2
\left(1-\frac{(P^2+p_1^2-p_2^2)^2}{4P^2p_1^2}\right)\left(1-\frac{(P^2+p_4^2-p_3^2)^2}{4P^2p_4^2}\right)
\Bigg]
\Bigg\},
\end{eqnarray}
Finally, in order to integrating over $\Omega_4$, we use the coordinate system with the z-axis along $\up_3$. 
Then, we get
\begin{equation}
K_3= \frac{\pi^3}{32 p_1^2 p_2^2 p_3^4 p_4^4} \int_{p_{34}}^{P_{34}} h(P)dP,
\end{equation}
where 
\begin{eqnarray}
h(P) &=& (P_{12}^2-P^2)(P_{34}^2-P^2)
\Theta(P_{12}-P)\Theta(P-p_{12})\nonumber \\ 
&\times&\Bigg\{ \left(P_{14}^2-P_{23}^2\right)^2
\Bigg[2\left(1+\frac{(P^2+p_1^2-p_2^2)(P^2+p_4^2-p_3^2)}{4P^2p_1 p_4}\right)^2
+\left(1-\frac{(P^2+p_1^2-p_2^2)^2}{4P^2p_1^2}\right)\left(1-\frac{(P^2+p_4^2-p_3^2)^2}{4P^2p_4^2}\right)\Bigg]
\nonumber \\ 
&-&4p_1p_4\left(P_{14}^2-P_{23}^2\right)\left(1+\frac{(P^2+p_1^2-p_2^2)(P^2+p_4^2-p_3^2)}{4P^2p_1 p_4}\right)\nonumber \\ 
&& \times\Bigg[2\left(1+\frac{(P^2+p_1^2-p_2^2)(P^2+p_4^2-p_3^2)}{4P^2p_1 p_4}\right)^2
+3\left(1-\frac{(P^2+p_1^2-p_2^2)^2}{4P^2p_1^2}\right)\left(1-\frac{(P^2+p_4^2-p_3^2)^2}{4P^2p_4^2}\right)\Bigg]
\nonumber \\ 
&+&p_1^2 p_4^2
\Bigg[8\left(1+\frac{(P^2+p_1^2-p_2^2)(P^2+p_4^2-p_3^2)}{4P^2p_1 p_4}\right)^4
+3\left(1-\frac{(P^2+p_1^2-p_2^2)^2}{4P^2p_1^2}\right)^2\left(1-\frac{(P^2+p_4^2-p_3^2)^2}{4P^2p_4^2}\right)^2\nonumber \\ 
&&+24 \left(1+\frac{(P^2+p_1^2-p_2^2)(P^2+p_4^2-p_3^2)}{4P^2p_1 p_4}\right)^2
\left(1-\frac{(P^2+p_1^2-p_2^2)^2}{4P^2p_1^2}\right)\left(1-\frac{(P^2+p_4^2-p_3^2)^2}{4P^2p_4^2}\right)
\Bigg]
\Bigg\}.
\end{eqnarray}
The result can be given in the same form as the other integrals in the previous subsections, i.e.,
\begin{eqnarray}
K_3= \frac{\pi^3}{32\,p_1^2\,p_2^2\,p_3^4\,p_4^4}\,L_3(p_{12},P_{12},p_{34},P_{34})\,,
\label{K3-intermed}
\end{eqnarray}
where
\begin{eqnarray}
L_3(a,b,c,d)&\equiv& \Theta(c-a)\Theta(d-b)\Theta(b-c)
J_3(c,b,b,d) +\Theta(a-c)\Theta(d-b)
J_3(a,b,b,d) \nonumber \\ 
&+&\Theta(a-c)\Theta(b-d)\Theta(d-a)
J_3(a,d,b,d) + \Theta(c-a)\Theta(b-d)
J_3(c,d,b,d)\,,
\end{eqnarray}
and
\begin{eqnarray}\hspace{-.2cm}
J_3(a,b,c,d)&=& \int_a^b dP\,(c^2-P^2)(d^2-P^2)\Bigg\{ \left(P_{14}^2-P_{23}^2\right)^2 \nonumber \\ 
&&\times
\Bigg[2\left(1+\frac{(P^2+p_1^2-p_2^2)(P^2+p_4^2-p_3^2)}{4P^2p_1 p_4}\right)^2
+\left(1-\frac{(P^2+p_1^2-p_2^2)^2}{4P^2p_1^2}\right)\left(1-\frac{(P^2+p_4^2-p_3^2)^2}{4P^2p_4^2}\right)\Bigg]
\nonumber \\ 
&-&4p_1p_4\left(P_{14}^2-P_{23}^2\right)\left(1+\frac{(P^2+p_1^2-p_2^2)(P^2+p_4^2-p_3^2)}{4P^2p_1 p_4}\right)\nonumber \\ 
&& \times\Bigg[2\left(1+\frac{(P^2+p_1^2-p_2^2)(P^2+p_4^2-p_3^2)}{4P^2p_1 p_4}\right)^2
+3\left(1-\frac{(P^2+p_1^2-p_2^2)^2}{4P^2p_1^2}\right)\left(1-\frac{(P^2+p_4^2-p_3^2)^2}{4P^2p_4^2}\right)\Bigg]
\nonumber \\ 
&+&p_1^2 p_4^2
\Bigg[8\left(1+\frac{(P^2+p_1^2-p_2^2)(P^2+p_4^2-p_3^2)}{4P^2p_1 p_4}\right)^4
+3\left(1-\frac{(P^2+p_1^2-p_2^2)^2}{4P^2p_1^2}\right)^2\left(1-\frac{(P^2+p_4^2-p_3^2)^2}{4P^2p_4^2}\right)^2\nonumber \\ 
&&+24 \left(1+\frac{(P^2+p_1^2-p_2^2)(P^2+p_4^2-p_3^2)}{4P^2p_1 p_4}\right)^2
\left(1-\frac{(P^2+p_1^2-p_2^2)^2}{4P^2p_1^2}\right)\left(1-\frac{(P^2+p_4^2-p_3^2)^2}{4P^2p_4^2}\right)
\Bigg]
\Bigg\}.
\end{eqnarray}
\end{widetext}
To leading order in powers of large $\mu$, the result reduces to
\begin{equation}
L_3(0,2\mu,0,2\mu) = \frac{29\times 2^{20}\mu^9}{45045} .
\end{equation}
Then, by making use of the relation in Eq.~(\ref{K3-intermed}), we derive
\begin{equation}
K_3 \simeq \frac{\pi^3}{2\mu^{12}}L_3(0,2\mu,0,2\mu) = \frac{29\times 2^{15}\pi^3}{45045 \mu^3}.
\label{I3}
\end{equation}

\subsection{$F_{r_1 r_2 r_3 r_4}$ to leading order in inverse powers of $\mu$}

By making use of the leading order results for $K_i$ ($i=1,2,3$) obtained in the previous subsections,
here we write down the explicit results for the functions $F_{r_1 r_2 r_3 r_4}$ in the same approximation:
\begin{subequations}
\begin{eqnarray}
F_{1111} &=&  \frac{1301\times  2^{10}\pi^{3}}{9009 \mu^3},
\end{eqnarray}
\begin{eqnarray}
\hspace*{-25pt}
F_{1221} &=&  F_{2112} = \frac{443 \times 2^{10}\pi^{3}}{9009 \mu^3},
\\  
\hspace*{-25pt}
F_{1112} &=& F_{1121} = F_{1211} = F_{2111}  = \frac{415 \times 2^{10}\pi^{3}}{9009 \mu^3},
\\  
\hspace*{-25pt}
F_{1122} &=&  F_{1212} = F_{2121} = F_{2211} = \frac{1357 \times 2^{10}\pi^{3}}{45045 \mu^3} , 
\\  
\hspace*{-25pt}
F_{1222} &=&  F_{2122} = F_{2212} = F_{2221} = \frac{359\times 2^{10}\pi^{3}}{45045 \mu^3}, 
\\  
\hspace*{-25pt}
F_{2222} &=& \frac{29\times 2^{15}\pi^{3}}{45045\mu^3}.
\end{eqnarray}
\end{subequations}
Note that 
\begin{equation}
\sum_{r_1,r_2,r_3,r_4} F_{r_1 r_2 r_3 r_4}= \frac{3\times 2^{10}\pi^{3}}{5\mu^3}.
\end{equation}

\end{document}